\documentclass[a4paper]{article}
\usepackage{amsmath}
\usepackage{amsfonts}
\usepackage{amssymb}
\usepackage{wasysym}
\usepackage{amsthm}
\usepackage{tikz-cd}
\usepackage{longtable}
\usepackage{empheq}
\usepackage{ytableau}
\usepackage{url}
\usepackage{dsfont}
\usepackage{cancel}
\usepackage{slashed}
\usepackage{scalerel}
\usepackage[pdftex]{pict2e}
\usepackage{scalerel}
\usepackage[dvipsnames]{xcolor}

\usepackage{authblk}

\newcommand\smallmath[2]{#1{\raisebox{\dimexpr \fontdimen 22 \textfont 2
      - \fontdimen 22 \scriptscriptfont 2 \relax}{$\scriptscriptstyle #2$}}}

\newcommand\smallotimes{\smallmath\mathbin\otimes}

\def\be{\begin{eqnarray}}
\def\ee{\end{eqnarray}}
\def\nn{\nonumber}

\def\ba{\begin{equation}\begin{aligned}}
\def\ea{\end{aligned}\end{equation}}

\def\horr{{{\smallsmile}\atop{\smallfrown}}}

\def\Ker{{\rm Ker}}
\def\Im{{\rm Im}}

\def\vth{\vartheta}

\def\cH{{\cal H}}
\def\cK{{\cal K}}
\def\cR{{\cal R}}

\newcommand{\verteq}{\rotatebox{90}{$\,=$}}
\newcommand{\vertLra}{\rotatebox{270}{$\,\Lra$}}
\newcommand{\equalto}[2]{\underset{\scriptstyle\overset{\mkern4mu\verteq}{#2}}{#1}}

\def\Lra{\Longrightarrow}

\usepackage[hidelinks,colorlinks=true,unicode]{hyperref}
\hypersetup{
	linkcolor=blue,
	citecolor=blue,
	filecolor=magenta,
	urlcolor=blue,
}
\usepackage[left=2cm,right=2cm,top=2cm,bottom=2cm,bindingoffset=0cm]{geometry}

\numberwithin{equation}{section}

\usepackage[natbib=true, backend = bibtex, style=numeric-comp, sorting=none,maxbibnames=99]{biblatex}
\date{}

\begin{document}
	
\title{\bf Khovanov complexes for bipartite links}

\author[2,3]{{\bf A. Anokhina}\thanks{\href{mailto:anokhina@itep.ru}{anokhina@itep.ru}}}
\author[1,2,3]{{\bf E. Lanina}\thanks{\href{mailto:lanina.en@phystech.edu}{lanina.en@phystech.edu}}}
\author[1,2,3]{{\bf A. Morozov}\thanks{\href{mailto:morozov@itep.ru}{morozov@itep.ru}}}

\vspace{5cm}

\affil[1]{Moscow Institute of Physics and Technology, 141700, Dolgoprudny, Russia}

\affil[2]{Institute for Information Transmission Problems, 127051, Moscow, Russia}
\affil[3]{NRC "Kurchatov Institute", 123182, Moscow, Russia\footnote{former Institute for Theoretical and Experimental Physics, 117218, Moscow, Russia}}

\renewcommand\Affilfont{\itshape\small}

\maketitle

\vspace{-7cm}

\begin{center}
	\hfill MIPT/TH-22/26\\
	\hfill ITEP/TH-26/26\\
	\hfill IITP/TH-24/26
\end{center}

\vspace{4.5cm}

\begin{abstract}
	{
Recently,  for a limited class for bipartite links, 
the complicated Khovanov--Rozansky matrix factorization technique was reduced to an
analogue of elementary Kauffman--Khovanov cycle calculus for an arbitrary $N$.
In this note, we demonstrate the consistency of such reduction  with
the computation of the bipartite Khovanov polynomials for $N=2$.
Namely, we explain how the Kauffman--Khovanov $2^2$-hypercube
is shrinked to the  bipartite 3-hypercube. 
    }
\end{abstract}

%\bigskip
%
%{\it ??????? It is expected to be read with Dima's text ``Notes on Khovanov-Rozansky construction''.}

\tableofcontents

%\section{Reduction of one bipartite vertex}

\section{Introduction}

In this paper, we show that a Khovanov 4-hypercube for bipartite links can be reduced to a 3-hypercube. This result is in full consistency with our recent works~\cite{ALM,ALM2,ALM3,2506.08721,2508.05191,lanina2026khovanov,galakhov2026reductions}.

We consider observables in the 3d Chern--Simons topological field theory~\cite{Chern-Simons} --- Wilson loops defined by a representation $R$ of the Lie algebra $\mathfrak{su}(N)$ and a knot $\cK$:
\begin{equation}
     \label{WilsonLoopExpValue}
         H_{R}^{\mathcal{K}}(q, A) = \left\langle \text{tr}_{R} \ P \exp \left( \oint_{\mathcal{K}} {\cal A} \right) \right\rangle_{\text{CS}},
         %\frac{1}{\dim_q(R)}
     \end{equation}
where the action is
\begin{equation}\label{CS-action}
         S_{\text{CS}}[{\cal A}] = \frac{\kappa}{4 \pi} \int_{S^3} \text{tr} \left( {\cal A} \wedge d{\cal A} +  \frac{2}{3} {\cal A} \wedge {\cal A} \wedge {\cal A} \right).
     \end{equation}
%In formula~\eqref{WilsonLoopExpValue}, $\dim_q(R)$ is the quantum dimension of a representation $R$. 
In a proper normalization, the Wilson loop~\eqref{WilsonLoopExpValue} turns out to be a polynomial in the variables $q = \exp\left(\frac{2\pi i}{\kappa +N}\right)$ and $A=q^N$. This construction is straightforwardly generalized to a link case. Due to the topological nature of the theory~\eqref{CS-action}, the Wilson loops~\eqref{WilsonLoopExpValue} are quantum knot invariants~\cite{Witten} called the colored (by a representation $R$) HOMFLY polynomials~\cite{freyd1985new,przytycki1988invariants}. One of the main features of the Chern--Simons theory is the fact that the observables~\eqref{WilsonLoopExpValue} can be calculated non-perturbatively via the so-called Reshetikhin--Turaev approach~\cite{reshetikhin1990ribbon,reshetikhin1991invariants,turaev1990yang}. The answer is expressed as the quantum trace of products of $\cR$-matrices of the quantized universal enveloping algebra $U_q(\mathfrak{sl}_N)$ associated of each crossing of a knot $\cK$.

In this text, we mainly consider the fundamental representation of the $SU(2)$ gauge group. In this case, the quantum knot invariants~\eqref{WilsonLoopExpValue} are the celebrated Jones polynomials~\cite{jones1985polynomial,jones1987hecke,jones2005jones}, and the $\cR$-matrix is very simple and given by the following expression:
\be
{\cal R}^{ij}_{kl} \ \stackrel{N=2}{=} \  \epsilon^{ij}\epsilon_{kl} -q\cdot \delta^i_k \delta^j_l\,,
\label{Krule}
\ee
see also Fig.\,\ref{fig:Kauff}. Thus, the $\cR$-matrix is decomposed into 2 resolutions, and the Jones polynomial can be graphically calculated via a 2-hypercube of resolutions, see examples in Figs.\,\ref{fig:Hopf-hypercube},\,\ref{fig:trefoil-hypercube}. If one considers the connection of 2 vertices, then one initially has 4 resolutions, and deals with a hypercube of $2^{2n}$ vertices, where $n$ is the number of double-crossings in a link. However, it is easy to show that this $2^{2n}$-hypercube can be reduced to a $3^n$-hypercube, see Fig.\,\ref{fig:pladeco-3-hyp}. For antiparallel connection of 2 vertices, we have also shown that one can draw such hypercubes already for the case of an arbitrary $N$~\cite{ALM}, and the analogous construction also works for the quantum knot invariants colored by (anti-)symmetric representations~\cite{ALM2}. 

The question is if this reduction works in the $T$-deformation of the Jones polynomial --- the Khovanov polynomial~\cite{khovanov2000categorification} believed to be the observable in the refined Chern--Simons theory~\cite{aganagic2015knot}. In this paper, we show that the answer is yes, and this result is in a full agreement with our works~\cite{2506.08721,2508.05191,lanina2026khovanov,galakhov2026reductions}.

\bigskip

\noindent The paper is organized as follows. In Section~\ref{sec:prelim}, we provide constructions of the Jones and Khovanov polynomials and also show that Jones hypercubes can be reduced in the case of links constructed as a connection of double vertices. Then, in Section~\ref{sec:KhRed}, we illustrate that the same reduction actually works in the $T$-deformed case. This result approves our constructions of~\cite{2506.08721,2508.05191,lanina2026khovanov,galakhov2026reductions} for a generic $N$ for bipartite links. Finally, we conclude in Section~\ref{sec:concl}.

\section{Preliminaries}\label{sec:prelim}

In this section, we introduce the Jones and Khovanov polynomials and explain the notions of 2- and 3-hypercubes.

\subsection{Jones polynomial}\label{sec:Jones}

\begin{figure}[h!]
\begin{picture}(300,250)(-80,-150)

\put(-60,0){\mbox{\bf J}}

\put(0,0){\mbox{$[\bigcirc]_{[0000]}^{\scaleto{\{0\}}{5.5pt}}$}}

\put(50,20){\mbox{$[\bigcirc\bigcirc]_{[0100]}^{\scaleto{\{1\}}{5.5pt}}$}}

\put(50,60){\mbox{$[\bigcirc\bigcirc]_{[0001]}^{\scaleto{\{1\}}{5.5pt}}$}}

\put(50,-20){\mbox{$[\bigcirc\bigcirc]_{[0010]}^{\scaleto{\{1\}}{5.5pt}}$}}

\put(50,-60){\mbox{$[\bigcirc\bigcirc]_{[1000]}^{\scaleto{\{1\}}{5.5pt}}$}}

\put(130,0){\mbox{$[\bigcirc]_{[1001]}^{\scaleto{\{2\}}{5.5pt}}$}}

\put(130,60){\mbox{$[\bigcirc]_{[0101]}^{\scaleto{\{2\}}{5.5pt}}$}}

\put(120,100){\mbox{$[\bigcirc\bigcirc\bigcirc]_{[0011]}^{\scaleto{\{2\}}{5.5pt}}$}}

\put(130,-40){\mbox{$[\bigcirc]_{[0110]}^{\scaleto{\{2\}}{5.5pt}}$}}

\put(123,-80){\mbox{$[\bigcirc\bigcirc\bigcirc]_{[1100]}^{\scaleto{\{2\}}{5.5pt}}$}}

\put(130,-120){\mbox{$[\bigcirc]_{[1010]}^{\scaleto{\{2\}}{5.5pt}}$}}

\put(210,35){\mbox{$[\bigcirc\bigcirc]_{[1011]}^{\scaleto{\{3\}}{5.5pt}}$}}

\put(210,100){\mbox{$[\bigcirc\bigcirc]_{[0111]}^{\scaleto{\{3\}}{5.5pt}}$}}

\put(210,-35){\mbox{$[\bigcirc\bigcirc]_{[1101]}^{\scaleto{\{3\}}{5.5pt}}$}}

\put(210,-100){\mbox{$[\bigcirc\bigcirc]_{[1110]}^{\scaleto{\{3\}}{5.5pt}}$}}

\put(325,0){\mbox{$[\bigcirc\bigcirc\bigcirc]_{[1111]}^{\scaleto{\{4\}}{5.5pt}}$}}

\put(20,12){\vector(1,1){40}}

\put(20,-9){\vector(1,-1){40}}

\put(90,70){\vector(1,1){25}}

\put(100,60){\vector(1,0){20}}

\put(90,50){\vector(1,-1.5){27}}

\put(85,10){\vector(1,-2){40}}

\put(90,-10){\vector(1,2){50}}

\put(80,-70){\vector(1,-1){40}}

\put(100,-65){\vector(2,-1){20}}

\put(83,-47){\vector(1,1){40}}

\put(185,100){\vector(1,0){20}}

\put(175,90){\vector(1,-1){45}}

\put(180,50){\vector(1,-2){37}}

\put(170,12){\vector(1.5,1){33}}

\put(170,-10){\vector(1.5,-1){33}}

\put(185,-85){\vector(2,-1){20}}

\put(176,-70){\vector(1,1){30}}

\put(180,-105){\vector(1,3){44}}

\put(262,35){\vector(2,-1){60}}

\put(262,-33){\vector(2,1){60}}

\put(25,10){\vector(2,1){20}}

\put(25,-8){\vector(2,-1){20}}

\put(90,30){\vector(1.5,1){30}}
\put(90,-30){\vector(1,-2){40}}
\put(180,-30){\vector(1,4){30}}
\put(180,-48){\vector(1,-1.5){27}}
\put(260,90){\vector(1,-1){80}}
\put(260,-93){\vector(1,1){80}}

\put(90,10){\vector(1,-1.5){27}}

\put(90,-28){\vector(3,-1){25}}

\put(180,70){\vector(1,1){25}}

\put(180,-110){\vector(2,1){23}}

\put(-70,-145){\mbox{$J^{3_1} \overset{\eqref{J-hypercube}}{=} q^{4}\cdot \ \ \ \ (D_2 \quad \quad \; - 4q D_2^2 \quad \quad \quad + 4q^2 D_2 + 2q^2 D_2^3 \quad \; \; - 3 q^3 D_2^2 \qquad \qquad \qquad \quad \quad \ + q^4 D_2^3 )$}}
    
\end{picture}
    \caption{\footnotesize Jones $2^4$-hypercube for the twist trefoil knot, see Fig.\,\ref{fig:bip-vert-links}. It can be reduced to $3^2$-hypercube, see Fig.\,\ref{fig:tw-trefoil-J-complex}.}
    \label{fig:4-trefoil-hypercube}
\end{figure}
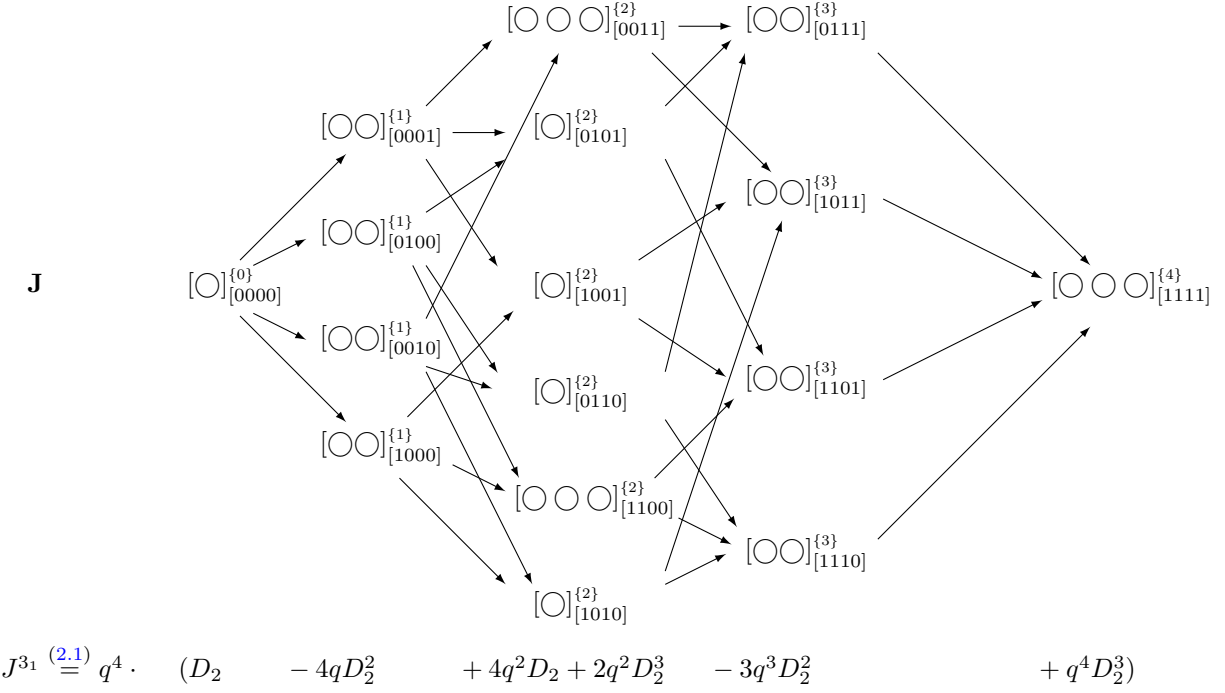

The Jones polynomial is defined by the Kauffman bracket~\cite{Kauff} in Fig.\,\ref{fig:Kauff} together with the requirement that each closed cycle contributes $D_2 = q + q^{-1}$ into the Jones polynomial. Each crossing is resolved in two ways, thus, resolutions of the whole link can be organised in a 2-hypercube, see examples in Figs.\,\ref{fig:Hopf-hypercube},\,\ref{fig:trefoil-hypercube},\,\ref{fig:4-trefoil-hypercube}. Different resolutions $r$ of a link stand in
vertices and are enumerated by words of zeros and unities $[\alpha_1 \ldots \alpha_n]$ with $\alpha_i = 0,\, 1$. 
Each resolution is associated with a set of $\nu(r)$ closed non-intersecting planar circles,
and these are the patterns which we associate with vertices. Actually, each circle consists of several segments between intersection points,
i.e. has a definite ``length''. The set of $\nu(r)$ lengths $l_a(r)$ can be also associated with the
vertex $r$ of the hypercube. Edges connect resolutions corresponding to elementary flips in which exactly one resolution of type $0$ is replaced by the resolution of type $1$. Resolutions of a link sharing the same vertical possess the same {\it height} $h := \sum_i \alpha_i$. Then, the Jones polynomial is
% which equals to
\begin{equation}\label{J-hypercube}
    J^{\cal L}(q) = (-)^{n_\circ} q^{n_\bullet - 2 n_\circ} \sum_{r}^{2^n} (-q)^{h(r)} D_2^{\nu(r)}\,.
\end{equation}
Here, the multiplier $(-)^{n_\circ} q^{n_\bullet - 2 n_\circ}$ is present to restore the topological invariance; $n_\circ$ and $n_\bullet$ are numbers of vertices depicted in Fig.\,\ref{fig:pos-neg-cr}. 

\begin{figure}[h!]
\begin{picture}(100,200)(-200,-100)

\put(0,70){

\put(-60,-20){\line(1,1){40}}
\put(-20,-20){\line(-1,1){18}}
\put(-60,20){\line(1,-1){18}}

\put(10,-2){\mbox{$=$}}

\qbezier(30,20)(50,0)(70,20)
\qbezier(30,-20)(50,0)(70,-20)

\put(85,-2){\mbox{$- \ \ \ q $}}

\qbezier(125,20)(145,0)(125,-20)
\qbezier(150,20)(130,0)(150,-20)

\put(-68,22){\mbox{$i$}}  \put(-15,22){\mbox{$j$}}  \put(-68,-28){\mbox{$k$}}  \put(-15,-28){\mbox{$l$}}

\put(35,-45){\mbox{type $0$}}

\put(125,-45){\mbox{type $1$}}
}

\put(0,-20){

\put(-60,-20){\line(1,1){18}}
\put(-20,-20){\line(-1,1){40}}
\put(-37.5,2){\line(1,1){18}}

\put(-68,22){\mbox{$i$}}  \put(-15,22){\mbox{$j$}}  \put(-68,-28){\mbox{$k$}}  \put(-15,-28){\mbox{$l$}}

\put(10,-2){\mbox{$=$}}

\qbezier(30,20)(50,0)(70,20)
\qbezier(30,-20)(50,0)(70,-20)

\put(85,-2){\mbox{$- \ \ \ q^{-1} $}}

\qbezier(125,20)(145,0)(125,-20)
\qbezier(150,20)(130,0)(150,-20)

\put(35,-45){\mbox{type $1$}}

\put(125,-45){\mbox{type $0$}}
}

\put(60,-20){

\put(-100,-70){\circle{30}}
\put(-75,-72.5){\mbox{$=\ \ \ D_2 \ = \ q+q^{-1}$}}
}

\end{picture}
\caption{\footnotesize The  celebrated Kauffman bracket --- the planar decomposition
of the ${\cal R}$-matrix vertex for the fundamental representation of $U_q(\mathfrak{sl}_2)$~\eqref{Krule}.
In this case ($N=2$), the conjugate of the fundamental representation is isomorphic to it,
thus, tangles in the picture have no orientation. The resolutions are of two types which we denote 0 and 1. These numbers are values of $\alpha_i$. 
}
\label{fig:Kauff}
\end{figure}

\begin{figure}[h!]
\centering
\begin{picture}(100,120)(-70,-15)

\put(-240,45){\mbox{\bf J}}

\put(-80,20){

\put(-100,25){\line(1,1){16}}
\put(-100,41){\line(1,-1){6}}
\put(-90,31){\line(1,-1){6}}

\put(34,0){

\put(-100,25){\line(1,1){16}}
\put(-100,41){\line(1,-1){6}}
\put(-90,31){\line(1,-1){6}}
}

% \put(-50,25){\line(-1,1){16}}
% \put(-50,41){\line(-1,-1){6}}
% \put(-60,31){\line(-1,-1){6}}

\put(-75,25){\oval(18,10)[b]}
\put(-75,41){\oval(18,10)[t]}

\put(-75,25){\oval(50,20)[b]}
\put(-75,41){\oval(50,20)[t]}
}

\put(-80,50){\mbox{$[\overset{2}{\bigcirc}\overset{2'}{\bigcirc}]_{[00]}^{\scaleto{\{0\}}{5.5pt}}$}}
\put(-35,65){\vector(1,1){20}}
\put(-35,40){\vector(1,-1){20}}
% \put(-40,75){\mbox{$m$}}
% \put(-40,25){\mbox{$m$}}

\put(-8,90){\mbox{$[\overset{4'}{\bigcirc}]_{[01]}^{\scaleto{\{1\}}{5.5pt}}$}}
\put(-8,10){\mbox{$[\overset{4}{\bigcirc}]_{[10]}^{\scaleto{\{1\}}{5.5pt}}$}}
% \put(-3,50){\mbox{$\bigoplus$}}

\put(-5,0){

\put(65,50){\mbox{$[\overset{\bar 2}{\bigcirc}\overset{\bar 2'}{\bigcirc}]_{[11]}^{\scaleto{\{2\}}{5.5pt}}$}}
\put(32,83){\vector(1,-1){27}}
\put(32,20){\vector(1,1){27}}
% \put(30,60){\mbox{$\Delta$}}
% \put(25,35){\mbox{$-\Delta$}}
}

\put(-150,-15){\mbox{$J^{\rm Hopf} \overset{\eqref{J-hypercube}}{=} q^{2}\cdot \ \ \ \ (D_2^2 \quad \quad \quad \ - 2q D_2 \quad \quad \quad \quad + q^2 D_2^2)$}}
    
\end{picture}
\caption{\footnotesize The Hopf link and its hypercube of resolutions.}
\label{fig:Hopf-hypercube}
\end{figure}

\begin{figure}[h!]
    \centering
\begin{picture}(300,140)(20,-25)

\put(-45,45){\mbox{\bf J}}

\put(90,45){\mbox{$[\overset{2}{\bigcirc}\overset{2'}{\bigcirc}\overset{2''}{\bigcirc}]_{[000]}^{\scaleto{\{0\}}{5.5pt}}$}}

\put(19,0){

\put(163,93){\mbox{$[\overset{4}{\bigcirc}\overset{2'}{\bigcirc}]_{[001]}^{\scaleto{\{1\}}{5.5pt}}$}}

\put(163,45){\mbox{$[\overset{2}{\bigcirc}\overset{4'}{\bigcirc}]_{[010]}^{\scaleto{\{1\}}{5.5pt}}$}}

\put(163,0){\mbox{$[\overset{4''}{\bigcirc}\overset{2''}{\bigcirc}]_{[100]}^{\scaleto{\{1\}}{5.5pt}}$}}

\put(130,57){\vector(1,1){30}}

\put(133,47){\vector(1,0){25}}

\put(130,38){\vector(1,-1){30}}

\put(17,0){

\put(225,93){\mbox{$[\overset{6}{\bigcirc}]_{[011]}^{\scaleto{\{2\}}{5.5pt}}$}}

\put(225,45){\mbox{$[\overset{6'}{\bigcirc}]_{[101]}^{\scaleto{\{2\}}{5.5pt}}$}}

\put(225,0){\mbox{$[\overset{6''}{\bigcirc}]_{[110]}^{\scaleto{\{2\}}{5.5pt}}$}}

\put(192,57){\vector(1,1){30}}

\put(192,10){\vector(1,1){30}}

\put(192,38){\vector(1,-1){30}}

\put(192,85){\vector(1,-1){30}}

\put(194,95){\vector(1,0){25}}

\put(194,2){\vector(1,0){25}}

\put(255,10){\vector(1,1){30}}

\put(255,85){\vector(1,-1){30}}

\put(290,45){\mbox{$[\overset{3}{\bigcirc}\overset{3'}{\bigcirc}]_{[111]}^{\scaleto{\{3\}}{5.5pt}}$}}

\put(260,47){\vector(1,0){25}}

}

}

\put(50,-25){\mbox{$J^{3_1} \overset{\eqref{J-hypercube}}{=} q^{3}\cdot \ \ \ \ (D_2^3 \quad \quad \quad \ - 3q D_2^2 \quad \quad \quad \quad + 3q^2 D_2 \quad \quad \quad \ -q^3 D_2^2)$}}

\begin{tikzpicture}[scale=0.4]

\draw (0,0) -- (1,1); % Draws a line from (0,0) to (4,4)
\draw (1,0) -- (0.65,0.35);
\draw (0,1) -- (0.35,0.65);

\draw (0,-2) -- (1,-1);
\draw (1,-2) -- (0.65,-1.65);
\draw (0,-1) -- (0.35,-1.35);

\draw (0,-4) -- (1,-3);
\draw (1,-4) -- (0.65,-3.65);
\draw (0,-3) -- (0.35,-3.35);

\draw (1,1) .. controls (3,3) and (3,-6) .. (1,-4);
\draw (1,0) .. controls (1.5,-0.5) .. (1,-1);
\draw (1,-2) .. controls (1.5,-2.5) .. (1,-3);

\draw (0,1) .. controls (-2,3) and (-2,-6) .. (0,-4);
\draw (0,0) .. controls (-0.5,-0.5) .. (0,-1);
\draw (0,-2) .. controls (-0.5,-2.5) .. (0,-3);

\end{tikzpicture}
\end{picture}
    \caption{\footnotesize The torus trefoil knot $3_1$ and its hypercube of resolutions.}
    \label{fig:trefoil-hypercube}
\end{figure}

\begin{figure}[h!]
    \centering
\begin{picture}(300,65)(-135,-35)

\put(-80,0){

\put(-60,-20){\vector(1,1){18}}
\put(-20,-20){\vector(-1,1){18}}
\put(-42.5,2.5){\vector(-1,1){18}}
\put(-37.5,2.5){\vector(1,1){18}}

\put(-40,0){\circle{6}}

\put(-5,-2){\mbox{$:=$}}

}

\put(-60,-20){\line(1,1){18}}
\put(-20,-20){\vector(-1,1){40}}
\put(-37.5,2.5){\vector(1,1){18}}

%\put(-45,-38){\mbox{\Large $+$}}

\put(90,0){

\put(-60,-20){\vector(1,1){18}}
\put(-20,-20){\vector(-1,1){18}}
\put(-42.5,2.5){\vector(-1,1){18}}
\put(-37.5,2.5){\vector(1,1){18}}

\put(-40,0){\circle*{6}}

\put(-5,-2){\mbox{$:=$}}

}

\put(170,0){

\put(-60,-20){\vector(1,1){40}}
\put(-20,-20){\line(-1,1){18}}
\put(-42.5,2.5){\vector(-1,1){18}}

%\put(-45,-38){\mbox{\Large $-$}}
}

\end{picture}
    \caption{\footnotesize Denotations for a crossing and its mirror.}
    \label{fig:pos-neg-cr}
\end{figure}

Let us now consider the connection of two vertices. We call the initial Jones hypercubes as $2^2$- or $2^{2n}$-hypercubes, where $n$ is the number of double vertices. If one applies the Kauffman decomposition twice, one obtains three resolutions, see Fig.\,\ref{fig:pladeco-3-hyp}. The examples of links built from double vertices in our schematic notations are present in Fig.\,\ref{fig:bip-vert-links}. Thus, for a link instead of $2^2$-hypercube, one can draw $3$-hypercube, see examples in Figs.\,\ref{fig:Hopf-J-complex},\,\ref{fig:tw-trefoil-J-complex}. The examples of explicit truncation of $4$-hypercubes to $3$-hypercubes are provided in Section~\ref{sec:trunc-hyp}. Such 3-hypercubes can be generalized to the HOMFLY (an arbitrary $N$) case for bipartite links~\cite{ALM}, see a brief review in Section~\ref{sec:bip-HOMFLY}.

\begin{figure}[h!]
\begin{picture}(100,120)(-90,-95)

\put(20,0){
\put(-105,15){\line(1,-1){12}} \put(-87,3){\line(1,1){12}}
\put(-93,-3){\line(-1,-1){12}} \put(-75,-15){\line(-1,1){12}}
\put(-90,0){\circle*{6}}  \put(-90,-9){\line(0,1){18}}

\put(-60,-2){\mbox{$:=$}}
}

\put(0,0){\put(-17,20){\line(1,-1){17}}\put(-17,20){\line(1,-1){14}}   \put(0,3){\line(1,1){17}}
 \put(0,-3){\line(-1,-1){17}}   \put(17,-20){\line(-1,1){17}} \put(17,-20){\line(-1,1){14}}
 \put(0,-3){\line(0,1){6}}}

\put(10,0){

\put(20,-2){\mbox{$:=$}}

\qbezier(50,20)(55,9)(58,4) \qbezier(63,-4)(85,-40)(110,20)
\qbezier(50,-20)(75,40)(97,4)  \qbezier(102,-4)(105,-9)(110,-20)

}

\put(20,0){
\put(120,-2){\mbox{$=$}}

\put(-25,0){

\put(100,65){
\put(70,-60){\line(1,0){40}}
\put(110,-70){\line(-1,0){40}}
\put(120,-67){\mbox{$-\ \ \ q$}}

\put(-40,0){
\put(210,-50){\line(0,-1){30}}
\put(220,-80){\line(0,1){30}}
}

\put(30,0){
\put(210,-50){\line(0,-1){30}}
\put(220,-80){\line(0,1){30}}
}

\put(190,-67){\mbox{$+\ \ \ q^{3}$}}

\put(75,-95){\mbox{type 0}}
\put(160,-95){\mbox{type 1}}
\put(230,-95){\mbox{type 2}}

}
}
}

\put(0,-65){

\put(20,0){
\put(-105,15){\line(1,-1){12}} \put(-87,3){\line(1,1){12}}
\put(-93,-3){\line(-1,-1){12}} \put(-75,-15){\line(-1,1){12}}
\put(-90,0){\circle{6}}  \put(-90,-9){\line(0,1){18}}

\put(-60,-2){\mbox{$:=$}}
}

\put(0,0){
\put(0,0){\put(-17,20){\line(1,-1){17}}\put(-17,20){\line(1,-1){14}}   \put(0,3){\line(1,1){17}}
 \put(0,-3){\line(-1,-1){17}}   \put(17,-20){\line(-1,1){17}} \put(17,-20){\line(-1,1){14}}
\put(-1,-3){\line(0,1){6}} \put(1,-3){\line(0,1){6}}
}

\put(30,-2){\mbox{$:=$}}

\put(10,0){
\qbezier(50,20)(75,-40)(97,-4)  \qbezier(102,4)(105,9)(110,20)
\qbezier(50,-20)(55,-9)(58,-4) \qbezier(63,4)(85,40)(110,-20)
}

\put(140,-2){\mbox{$=$}}

\put(0,0){

\put(100,65){
\put(70,-60){\line(1,0){40}}
\put(110,-70){\line(-1,0){40}}
\put(120,-67){\mbox{$-\ \ \ q^{-1}$}}

\put(-40,0){
\put(210,-50){\line(0,-1){30}}
\put(220,-80){\line(0,1){30}}
}

\put(30,0){
\put(210,-50){\line(0,-1){30}}
\put(220,-80){\line(0,1){30}}
}

\put(188,-67){\mbox{$+ \ \ q^{-3}$}}

\put(75,-95){\mbox{type 2}}
\put(160,-95){\mbox{type 1}}
\put(230,-95){\mbox{type 0}}

}}}

}

\end{picture}
\caption{\footnotesize
The Kauffman resolutions of the positive (in the first line) and negative (in the second line) double vertices.
}\label{fig:pladeco-3-hyp}
\end{figure}
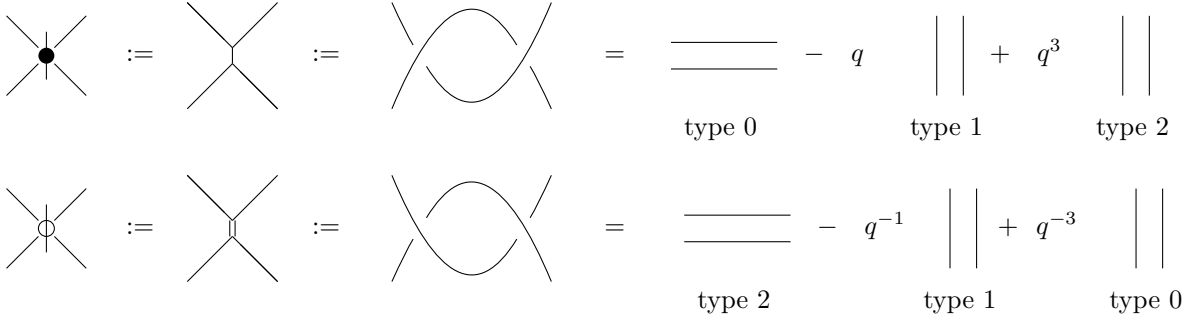

\begin{figure}[h!]
\begin{picture}(100,110)(-30,-60)

\put(0,0){

\put(20,0){

\put(-6,0){\line(1,0){13}}
\put(-23,17){\line(1,-1){17}}\put(-23,17){\vector(1,-1){14}}   \put(6,0){\vector(1,1){17}}
\put(-6,0){\vector(-1,-1){17}}   \put(23,-17){\line(-1,1){17}} \put(23,-17){\vector(-1,1){14}}

\qbezier(-23,17)(-43,45)(0,45) \qbezier(23,17)(43,45)(0,45)
\qbezier(-23,-17)(-43,-45)(0,-45) \qbezier(23,-17)(43,-45)(0,-45)
}

\put(130,0){

\put(-3,0){
\qbezier(-3,0)(-13,-7)(-12,-10) \put(-12,-10){\vector(0,-1){2}}
\qbezier(8,0)(18,-7)(16,-10) \put(10,-2){\vector(-1,1){2}}
\put(-3,0){\line(1,0){10}}

\put(-13,10){\line(1,-1){10}}
\put(9,0){\line(1,1){10}}
}

\put(-3,-20){
\qbezier(-3,0)(-13,7)(-12,10) \put(-6,2){\vector(1,-1){2}}
\qbezier(8,0)(18,7)(16,10) \put(16,9){\vector(0,1){2}}
\put(-3,0){\line(1,0){11}}
}

\put(-6,0){\line(1,0){13}}
\put(-23,17){\line(1,-1){17}}\put(-23,17){\vector(1,-1){14}}   \put(6,0){\vector(1,1){17}}

\put(0,-20){
\put(-6,0){\vector(-1,-1){17}}   \put(23,-17){\line(-1,1){17}} \put(23,-17){\vector(-1,1){14}}
}

\qbezier(-23,17)(-43,45)(0,45) \qbezier(23,17)(43,45)(0,45)

\put(0,-20){
\qbezier(-23,-17)(-43,-45)(0,-45) \qbezier(23,-17)(43,-45)(0,-45)
}
}

\put(250,0){  

\put(-6,0){\line(1,0){13}}
\put(-23,17){\line(1,-1){17}}
\put(-23,17){\vector(1,-1){14}}   
\put(6,0){\vector(1,1){17}}
\put(-6,0){\vector(-1,-1){17}}   \put(23,-17){\line(-1,1){17}} 
\put(23,-17){\vector(-1,1){14}}

\qbezier(-23,17)(-37,36)(-37,0) \qbezier(-23,-17)(-37,-36)(-37,0)
\qbezier(23,17)(37,36)(37,0) \qbezier(23,-17)(37,-36)(37,0)

}

\put(360,0){

\qbezier(17,10)(32,26)(40,19)

\qbezier(17,-10)(32,-26)(40,-19)

\qbezier(-13,10)(-21,20)(-13,30)

\qbezier(-13,-30)(-21,-20)(-13,-10)

\qbezier(60,19)(65,25)(60,30)

\qbezier(-13,30)(24,60)(60,30)

\qbezier(60,-30)(65,-24)(60,-19)

\qbezier(-13,-30)(24,-60)(60,-30)

\put(-13,10){\vector(1,-1){10}}

\put(7,0){\vector(1,1){10}}

\put(17,-10){\vector(-1,1){10}}

\put(-3,0){\vector(-1,-1){10}}

\put(-3,0){\line(1,0){10}}

\put(50,-9){\line(0,1){18}}

\put(60,-19){\vector(-1,1){10}}
\put(50,-9){\vector(-1,-1){10}}

\put(40,19){\vector(1,-1){10}}
\put(50,9){\vector(1,1){10}}

}
}

\end{picture}
    \caption{\footnotesize From left to right: the 2-unknot, the 4-unknot, the Hopf link and the twist trefoil knot $3_1$.}
    \label{fig:bip-vert-links}
\end{figure}
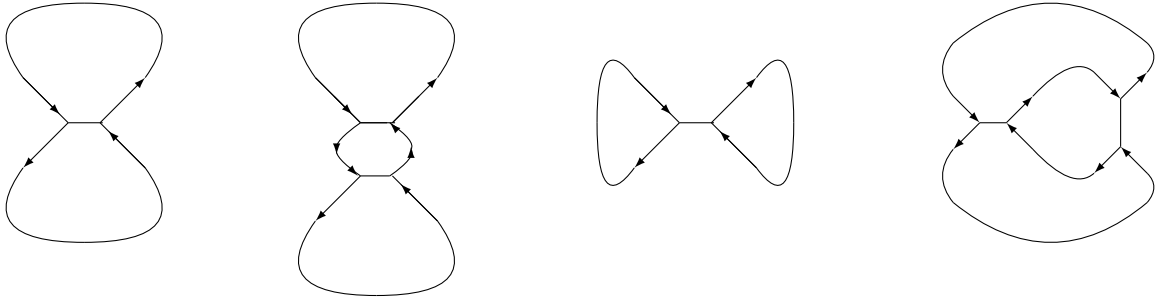

\begin{figure}[h!]
    \begin{equation}\nn
\begin{aligned}
    {\rm\bf J} \ \ \ \ \ \ \ \ \ \ \ \ \ \ \ &[\overset{1}{\bigcirc} \overset{1'}{\bigcirc}]_{[0]}^{\scaleto{\{0\}}{5.5pt}} \; \longrightarrow \; [\overset{2}{\bigcirc}]_{[1]}^{\scaleto{\{1\}}{5.5pt}} \; {\color{Green} \longrightarrow} \; [\overset{{\color{Green} 2}}{\bigcirc}]_{[2]}^{\scaleto{\{3\}}{5.5pt}} \\
    J^{\rm Hopf} = q^2 \cdot \ \ \ &(D_2^2 \quad \quad \quad \ -q D_2 \ \ \quad \ +q^3 D_2)
\end{aligned}
\end{equation}
    \caption{\footnotesize The 3-hypercube for the Hopf link. %We color in green the arrow for the sewing operator. The boxed elements connected by this blue arrow are shrinked when transferring to the 2-hypercube in Fig.\,\ref{fig:H-Hopf-complex}.
    } 
    \label{fig:Hopf-J-complex}
\end{figure}

\begin{figure}[h!]
\centering
\begin{picture}(100,200)(40,-55)

\put(-140,50){\mbox{\bf J}}

\put(5,0){
\put(-65,50){\mbox{$[\overset{4}{\bigcirc}]_{[00]}^{\scaleto{\{0\}}{5.5pt}}$}}
\put(-35,60){\vector(1,1){20}}
\put(-35,42){\vector(1,-1){20}}
}

\put(-10,90){\mbox{$[\overset{2}{\bigcirc}\overset{2'}{\bigcirc}]_{[01]}^{\scaleto{\{1\}}{5.5pt}}$}}
\put(-10,10){\mbox{$[\overset{\bar 2}{\bigcirc}\overset{\bar 2'}{\bigcirc}]_{[10]}^{\scaleto{\{1\}}{5.5pt}}$}}

\put(10,0){

\put(40,130){\mbox{$[\overset{{\color{blue} 2}}{\bigcirc}\overset{{\color{blue} 2'}}{\bigcirc}]_{[02]}^{\scaleto{\{3\}}{5.5pt}}$}}
\put(40,-30){\mbox{$[\overset{{\color{blue} \bar 2}}{\bigcirc}\overset{{\color{blue} \bar 2'}}{\bigcirc}]_{[20]}^{\scaleto{\{3\}}{5.5pt}}$}}
{\color{blue} 
\put(10,105){\vector(1,1){25}}
\put(10,0){\vector(1,-1){25}}}

\put(40,50){\mbox{$[\overset{4'}{\bigcirc}]_{[11]}^{\scaleto{\{2\}}{5.5pt}}$}}
\put(10,80){\vector(1,-1){25}}
\put(10,25){\vector(1,1){25}}

\put(10,0){

\put(70,125){\vector(1,-1){25}}
\put(70,-15){\vector(1,1){25}}
\put(100,90){\mbox{$[\overset{{\color{Green} 4}}{\bigcirc}]_{[12]}^{\scaleto{\{4\}}{5.5pt}}$}}
\put(100,10){\mbox{$[\overset{{\color{Green} 4'}}{\bigcirc}]_{[21]}^{\scaleto{\{4\}}{5.5pt}}$}}
{\color{Green} 
\put(68,63){\vector(1,1){27}}
\put(68,45){\vector(1,-1){27}}}

\put(-35,0){

{\color{Green} \put(150,80){\vector(1,-1){25}}
\put(150,25){\vector(1,1){25}}}
\put(180,50){\mbox{$[\bigcirc]_{[22]}^{\scaleto{\{6\}}{5.5pt}}$}}

}
}
}

\put(-110,-55){\mbox{$J^{3_1}= q^3\cdot \ \ (D_2 \ \quad \quad - 2qD_2^2 \ \ \ +q^2 D_2 + 2q^3 D_2^2 \; \ - 2q^4D_2 \; \ \ + q^6 D_2)$}}
    
\end{picture}
\caption{\footnotesize The 3-hypercube for the trefoil knot in the twist presentation, see Fig.\,\ref{fig:bip-vert-links}. Green arrows will be associated with the zero morphisms, and blue arrows will correspond to the Sh morphisms, see Section~\ref{sec:KhRed}. Spaces are enumerated by $[\alpha_1\, \alpha_2]$ with $\alpha_1=0,\,1,\,2$ corresponding to smoothings shown in Fig.\,\ref{fig:pladeco-3-hyp}.}
\label{fig:tw-trefoil-J-complex}
\end{figure}

\newpage

\subsection{Truncation of hypercubes}\label{sec:trunc-hyp}

For the Khovanov polynomial $(N=2)$, one can explicitly observe a reduction from a $2^{2n}$-hypercube to a $3^{n}$-hypercube, where $n$ is the number of double crossings in a link, what we will demonstrate in Section~\ref{sec:KhRed}. Such a phenomenon can be shown already at the level of the Jones polynomial~\cite{2506.08721}, see subsections below.   

\subsubsection{2-unknot}\label{sec:2-unknot-red}

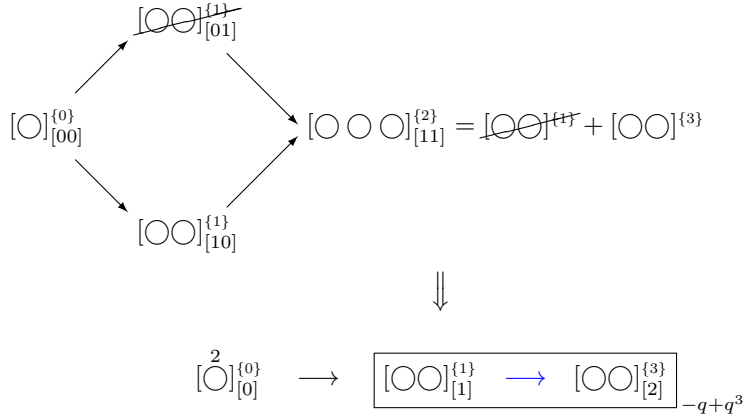
\begin{figure}[h!]
\centering
\begin{picture}(100,150)(50,-50)

%\put(-120,47.5){\mbox{\bf J}}

\put(-60,50){\mbox{$[\bigcirc]_{[00]}^{\scaleto{\{0\}}{5.5pt}}$}}
\put(-35,65){\vector(1,1){20}}
\put(-35,40){\vector(1,-1){20}}
% \put(-40,75){\mbox{$m$}}
% \put(-40,25){\mbox{$m$}}

\put(-12,90){\mbox{$\cancel{[\bigcirc\bigcirc]_{[01]}^{\scaleto{\{1\}}{5.5pt}}}$}}
\put(-12,10){\mbox{$[\bigcirc\bigcirc]_{[10]}^{\scaleto{\{1\}}{5.5pt}}$}}
% \put(-3,50){\mbox{$\bigoplus$}}

\put(-10,0){

\put(62,50){\mbox{$[\bigcirc\bigcirc\bigcirc]_{[11]}^{\scaleto{\{2\}}{5.5pt}}=\cancel{[\bigcirc\bigcirc]^{\scaleto{\{1\}}{5.5pt}}}+[\bigcirc\bigcirc]^{\scaleto{\{3\}}{5.5pt}}$}}
\put(32,80){\vector(1,-1){27}}
\put(32,22){\vector(1,1){27}}
% \put(30,60){\mbox{$\Delta$}}
% \put(25,35){\mbox{$-\Delta$}}
}

\put(0,-45){\mbox{$\quad[\overset{2}{\bigcirc}]_{[0]}^{\scaleto{\{0\}}{5.5pt}} \quad \longrightarrow \quad \boxed{[\bigcirc\bigcirc]_{[1]}^{\scaleto{\{1\}}{5.5pt}} \quad {\color{blue} \longrightarrow} \quad  [\bigcirc\bigcirc]_{[2]}^{\scaleto{\{3\}}{5.5pt}}}_{\,-q+q^3}$}}

\put(100,0){\mbox{$\vertLra$}}
    
\end{picture}
\caption{\footnotesize Truncation of the Jones $2^2$-hypercube of resolutions to 3-hypercube for 2-unknot. One cycle $\bigcirc$ giving contribution $D_2$ can be expanded as $D_2 = q + q^{-1}$, so that $\bigcirc^{\scaleto{\{0\}}{5.5pt}} = \{0\}^{\scaleto{\{-1\}}{5.5pt}} + \{0\}^{\scaleto{\{+1\}}{5.5pt}}$ in our cycle notation. Here, this substitution is made in the right corner. After that, one of the middle pair of cycles added to the right one vanishes because middle cycles carry the minus sign.}
\label{fig:2-unknot-hypercube}
\end{figure}

For the 2-unknot, we have $2^2$-hypercube, see Fig.\,\ref{fig:2-unknot-hypercube}. {\bf Kauffman calculation for two vertices and $N=2$} gives:
\be
J^{\text{2-unknot}}=\frac{1}{q^4}\left(D_2 -2q D_2^2 + q^2 D_2^3\right) = \frac{D_2}{q^4}\Big(1-2q(q+q^{-1}) + q^2(q+q^{-1})^2\Big)=D_2
\ee
where $D_2 =D_{N=2} = q+q^{-1}$.
Of interest for us will be reordering of the r.h.s.:
\ba
J^{\text{2-unknot}}&=\frac{D_2}{q^4}\left(
\begin{array}{rcl}
1&- q(q+q^{-1}) &+q^3(q+q^{-1}) \\
\hline
& - q(q+q^{-1})& + q(q+q^{-1})
\end{array} \right)
%= \nn \\
= \\ 
&=\frac{D_2}{q^4}\Big(1- q(q+q^{-1})\underbrace{- q(q+q^{-1}) + q(q+q^{-1})}_0 +q^3(q+q^{-1})\Big)\,.
\label{reordunknot1}
\ea
It can be considered as the truncation of the $2^2$-hypercube to $3^1$-hypercube. The first underlined row at the l.h.s. of (\ref{reordunknot1}) is exactly this expression at $N=2$,
while and in the second line we have a cancelation. %???which will be inherited by cohomology calculus below???.

% \noindent {\bf Bipartite calculation}:

% \begin{equation}
%     \bigcirc \quad --- \quad \phi \, \bigcirc\bigcirc
% \end{equation}

% \be
% H^{\text{2-unknot}} = \frac{1}{A^2}(D + \phi D^2) = \frac{D}{A^2}(1+\phi D)=D
% \ee

\newpage

\subsubsection{4-unknot}

\begin{figure}[h!]
\begin{picture}(300,480)(-20,-360)

\put(0,0){\mbox{$[\bigcirc]_{[0000]}^{\scaleto{\{0\}}{5.5pt}}$}}

\put(50,20){\mbox{$[\bigcirc\bigcirc]_{[0100]}^{\scaleto{\{1\}}{5.5pt}}$}}

\put(50,60){\mbox{$\cancel{[\bigcirc\bigcirc]_{[0001]}^{\scaleto{\{1\}}{5.5pt}}}$}}

\put(50,-20){\mbox{$[\bigcirc\bigcirc]_{[0010]}^{\scaleto{\{1\}}{5.5pt}}$}}

\put(50,-60){\mbox{$\cancel{[\bigcirc\bigcirc]_{[1000]}^{\scaleto{\{1\}}{5.5pt}}}$}}

\put(120,0){\mbox{$\cancel{[\bigcirc\bigcirc\bigcirc]_{[1001]}^{\scaleto{\{2\}}{5.5pt}}}$}}

\put(120,60){\mbox{$\equalto{[\bigcirc\bigcirc\bigcirc]_{[0101]}^{\scaleto{\{2\}}{5.5pt}}}{\cancel{[\bigcirc\bigcirc]^{\scaleto{\{1\}}{5pt}}}+[\bigcirc\bigcirc]^{\scaleto{\{3\}}{5pt}}}$}}

\put(120,100){\mbox{$\cancel{[\bigcirc\bigcirc\bigcirc]_{[0011]}^{\scaleto{\{2\}}{5.5pt}}}$}}

\put(120,-40){\mbox{$[\bigcirc\bigcirc\bigcirc]_{[0110]}^{\scaleto{\{2\}}{5.5pt}}$}}

\put(123,-80){\mbox{$\cancel{[\bigcirc\bigcirc\bigcirc]_{[1100]}^{\scaleto{\{2\}}{5.5pt}}}$}}

\put(120,-120){\mbox{$\equalto{[\bigcirc\bigcirc\bigcirc]_{[1010]}^{\scaleto{\{2\}}{5.5pt}}}{\cancel{[\bigcirc\bigcirc]^{\scaleto{\{1\}}{5pt}}}+[\bigcirc\bigcirc]^{\scaleto{\{3\}}{5pt}}}$}}

\put(200,35){\mbox{$\equalto{[\bigcirc\bigcirc\bigcirc\bigcirc]_{[1011]}^{\scaleto{\{3\}}{5.5pt}}}{\cancel{[\bigcirc\bigcirc\bigcirc]^{\scaleto{\{2\}}{5pt}}}+\cancel{[\bigcirc\bigcirc\bigcirc]^{\scaleto{\{4\}}{5pt}}}}$}}

\put(200,100){\mbox{$\equalto{[\bigcirc\bigcirc\bigcirc\bigcirc]_{[0111]}^{\scaleto{\{3\}}{5.5pt}}}{\cancel{[\bigcirc\bigcirc\bigcirc]^{\scaleto{\{2\}}{5pt}}}+[\bigcirc\bigcirc\bigcirc]^{\scaleto{\{4\}}{5pt}}}$}}

\put(200,-35){\mbox{$\equalto{[\bigcirc\bigcirc\bigcirc\bigcirc]_{[1101]}^{\scaleto{\{3\}}{5.5pt}}}{\cancel{[\bigcirc\bigcirc\bigcirc]^{\scaleto{\{2\}}{5pt}}}+\cancel{[\bigcirc\bigcirc\bigcirc]^{\scaleto{\{4\}}{5pt}}}}$}}

\put(200,-100){\mbox{$\equalto{[\bigcirc\bigcirc\bigcirc\bigcirc]_{[1110]}^{\scaleto{\{3\}}{5.5pt}}}{\cancel{[\bigcirc\bigcirc\bigcirc]^{\scaleto{\{2\}}{5pt}}}+[\bigcirc\bigcirc\bigcirc]^{\scaleto{\{4\}}{5pt}}}$}}

\put(320,0){\mbox{$\equalto{[\bigcirc\bigcirc\bigcirc\bigcirc\bigcirc]_{[1111]}^{\scaleto{\{4\}}{5.5pt}}}{\cancel{[\bigcirc\bigcirc\bigcirc]^{\scaleto{\{2\}}{5pt}}}+\cancel{2[\bigcirc\bigcirc\bigcirc]^{\scaleto{\{4\}}{5pt}}}+[\bigcirc\bigcirc\bigcirc]^{\scaleto{\{6\}}{5pt}}}$}}

\put(20,12){\vector(1,1){40}}

\put(20,-9){\vector(1,-1){40}}

\put(90,70){\vector(1,1){25}}

\put(100,60){\vector(1,0){20}}

\put(90,50){\vector(1,-1.5){27}}

\put(85,10){\vector(1,-2){40}}

\put(90,-10){\vector(1,2){50}}

\put(80,-70){\vector(1,-1){40}}

\put(100,-65){\vector(2,-1){20}}

\put(83,-47){\vector(1,1){40}}

\put(185,100){\vector(1,0){20}}

\put(175,90){\vector(1,-1){45}}

\put(180,50){\vector(1,-2){37}}

\put(170,12){\vector(1.5,1){33}}

\put(170,-10){\vector(1.5,-1){33}}

\put(185,-85){\vector(2,-1){20}}

\put(176,-70){\vector(1,1){30}}

\put(180,-105){\vector(1,3){44}}

\put(282,35){\vector(2,-1){60}}

\put(282,-33){\vector(2,1){60}}

\put(200,-150){\mbox{$\vertLra$}}

\put(130,-320){

\put(5,0){
\put(-65,50){\mbox{$[\bigcirc]_{[00]}^{\scaleto{\{0\}}{5.5pt}}$}}
\put(-35,60){\vector(1,1){20}}
\put(-35,42){\vector(1,-1){20}}
}

\put(-10,90){\mbox{$[\bigcirc\bigcirc]_{[01]}^{\scaleto{\{1\}}{5.5pt}}$}}
\put(-10,10){\mbox{$[\bigcirc\bigcirc]_{[10]}^{\scaleto{\{1\}}{5.5pt}}$}}

\put(10,0){

\put(43,130){\mbox{$[\bigcirc\bigcirc]_{[02]}^{\scaleto{\{3\}}{5.5pt}}$}}
\put(43,-30){\mbox{$[\bigcirc\bigcirc]_{[20]}^{\scaleto{\{3\}}{5.5pt}}$}}
{\color{blue} 
\put(10,105){\vector(1,1){25}}
\put(10,0){\vector(1,-1){25}}}

\put(40,50){\mbox{$[\bigcirc\bigcirc\bigcirc]_{[11]}^{\scaleto{\{2\}}{5.5pt}}$}}
\put(10,80){\vector(1,-1){25}}
\put(10,25){\vector(1,1){25}}

\put(20,0){

\put(70,125){\vector(1,-1){25}}
\put(70,-15){\vector(1,1){25}}
\put(100,90){\mbox{$[\bigcirc\bigcirc\bigcirc]_{[12]}^{\scaleto{\{4\}}{5.5pt}}$}}
\put(100,10){\mbox{$[\bigcirc\bigcirc\bigcirc]_{[21]}^{\scaleto{\{4\}}{5.5pt}}$}}
{\color{blue} 
\put(68,63){\vector(1,1){27}}
\put(69,40){\vector(1,-1){25}}}

\put(-30,0){

{\color{blue} \put(150,80){\vector(1,-1){25}}
\put(150,25){\vector(1,1){25}}}
\put(180,50){\mbox{$[\bigcirc\bigcirc\bigcirc]_{[22]}^{\scaleto{\{6\}}{5.5pt}}$}}

}
}
}

}

\thicklines

\put(25,10){\vector(2,1){20}}

\put(25,-8){\vector(2,-1){20}}

{\color{blue}
\put(90,30){\vector(1.5,1){30}}
\put(90,-30){\vector(1,-2){40}}
\put(180,-30){\vector(1,4){30}}
\put(180,-48){\vector(1,-1.5){27}}
\put(280,90){\vector(1,-1){80}}
\put(280,-93){\vector(1,1){60}}
}

\put(90,10){\vector(1,-1.5){27}}

\put(90,-28){\vector(3,-1){25}}

\put(180,70){\vector(1,1){25}}

\put(180,-110){\vector(2,1){23}}
    
\end{picture}
    \caption{\footnotesize Jones $2^4$-hypercube for the 4-unknot. It can be reduced to $3^2$-hypercube. In the above picture, the edges remaining after the reduction are thick. New enumerators $[\alpha_1^{(3)} \alpha_2^{(3)}]$ are expressed through old ones $[\alpha_1^{(2)}\alpha_2^{(2)}\alpha_3^{(2)}\alpha_4^{(2)}]$ in the following way: $\alpha_1^{(3)}=\alpha_1^{(2)}+\alpha_3^{(2)}$, $\alpha_2^{(3)}=\alpha_2^{(2)}+\alpha_4^{(2)}$. In this example, we again perform expansions $\bigcirc^{\scaleto{\{0\}}{5.5pt}} = \{0\}^{\scaleto{\{-1\}}{5.5pt}} + \{0\}^{\scaleto{\{+1\}}{5.5pt}}$ meaning $D_2 = q + q^{-1}$ substitution in terms of the Jones contribution. Then, some neighboring sets of cycles vanish and leave the $3$-hypercube as the result.}
    \label{fig:4-unknot-hypercube}
\end{figure}
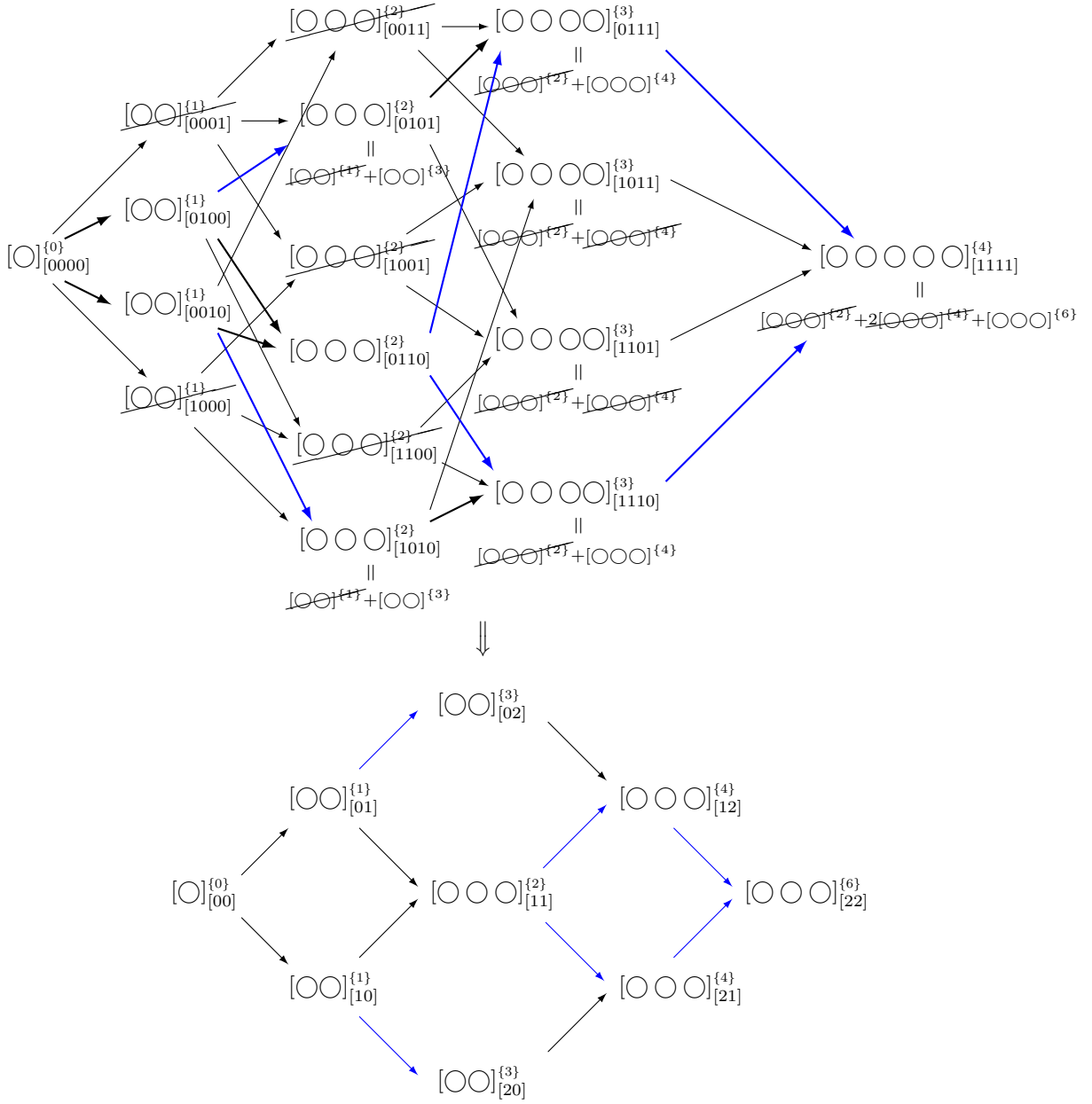

% \noindent {\bf Bipartite calculation}:
% \be
% \frac{1}{A^4}(D + 2\phi D^2+\phi^2 D^3) = \frac{D}{A^4}(1+\phi D)^2=D
% \ee

For the 4-unknot, we have $2^4$-hypercube, see Fig.\,\ref{fig:4-unknot-hypercube}. {\bf Kauffman calculation for four vertices and $N=2$} gives:

\ba
J^{\text{4-unknot}}&=\frac{D_2 -4qD_2^2 +6q^2D_2^3-4q^3D_2^4+ q^4D_2^5}{q^8}
= \\ 
&=\frac{D_2}{q^8}\Big(1-4q(q+q^{-1})+6q^2(q+q^{-1})^2-4q^3(q+q^{-1})^3 + q^4(q+q^{-1})^4\Big)=D_2\,.
\ea
It is a sum over $16$ vertices of a $2^4$-hypercube.
Of interest for us is the reordering:
\be
%??????\frac{{\bf D}}{q^8}\Big(1- q(q+q^{-1})\underbrace{- q(q+q^{-1}) + q(q+q^{-1})}_0 +q^3(q+q^{-1})\Big)
J^{\text{4-unknot}}=\frac{D_2}{q^8}\Big(1-4q(q+q^{-1})+6q^2(q+q^{-1})^2-4q^3(q+q^{-1})^3 + q^4(q+q^{-1})^4\Big)=
\ \ \ \ \ \ \ \ \ \ \ \ \ \ \ \ \ \
\nn \\
=\frac{D_2}{q^8}\left(
\begin{array}{rcccc}
1&-2q(q+q^{-1})&+2q^3(q+q^{-1}) &-2q^4(q+q^{-1})^2 &+(q^2 + q^6)(q+q^{-1})^2 \\
\hline
 &-2q(q+q^{-1})& +2q(q+q^{-1})   &     &  \\
 && + 4q^2(q+q^{-1})^2 &-4q^2(q+q^{-1})^2& \\
 &&& -2q^4(q+q^{-1})^2  & + 2q^4(q+q^{-1})^2
\end{array}
\right)=D_2\,,
\ee
where in the last three lines we observe full cancellations.
%-- which will be further promoted to the level of complexes.???
This can be considered as truncation of the $2^4$-hypercube to $3^2$-hypercube, see Fig.\ref{fig:4-unknot-hypercube}.

\subsection{Bipartite HOMFLY from hypercube}\label{sec:bip-HOMFLY}

\begin{figure}[h!]
\begin{picture}(100,110)(-110,-85)

\put(20,0){
\put(-105,15){\vector(1,-1){12}} \put(-87,3){\vector(1,1){12}}
\put(-93,-3){\vector(-1,-1){12}} \put(-75,-15){\vector(-1,1){12}}
\put(-90,0){\circle*{6}}  \put(-90,-9){\line(0,1){18}}

\put(-60,-2){\mbox{$:=$}}
}

\put(0,0){\put(-17,20){\line(1,-1){17}}\put(-17,20){\vector(1,-1){14}}   \put(0,3){\vector(1,1){17}}
 \put(0,-3){\vector(-1,-1){17}}   \put(17,-20){\line(-1,1){17}} \put(17,-20){\vector(-1,1){14}}
 \put(0,-3){\line(0,1){6}}}

\put(10,0){

\put(20,-2){\mbox{$:=$}}

\qbezier(50,20)(55,9)(58,4) \qbezier(63,-4)(85,-40)(110,20)
\put(56,8){\vector(1,-2){2}} \put(90,-13){\vector(1,1){2}} \put(109,18){\vector(1,2){2}}
\qbezier(50,-20)(75,40)(97,4)  \qbezier(102,-4)(105,-9)(110,-20)
\put(104,-8){\vector(-1,2){2}} \put(70,13){\vector(-1,-1){2}} \put(51,-18){\vector(-1,-2){2}}

}

\put(20,0){
\put(120,-2){\mbox{$=$}}

\put(-25,0){

\put(100,65){
\put(70,-60){\vector(1,0){40}}
\put(110,-70){\vector(-1,0){40}}
\put(120,-67){\mbox{$+\ \ \ (q^{N+1}-q^{N-1})$}}
\put(210,-50){\vector(0,-1){30}}
\put(220,-80){\vector(0,1){30}}
}
}
}

\put(0,-60){

\put(20,0){
\put(-105,15){\vector(1,-1){12}} \put(-87,3){\vector(1,1){12}}
\put(-93,-3){\vector(-1,-1){12}} \put(-75,-15){\vector(-1,1){12}}
\put(-90,0){\circle{6}}  \put(-90,-9){\line(0,1){18}}

\put(-60,-2){\mbox{$:=$}}
}

\put(0,0){
\put(0,0){\put(-17,20){\line(1,-1){17}}\put(-17,20){\vector(1,-1){14}}   \put(0,3){\vector(1,1){17}}
 \put(0,-3){\vector(-1,-1){17}}   \put(17,-20){\line(-1,1){17}} \put(17,-20){\vector(-1,1){14}}
\put(-1,-3){\line(0,1){6}} \put(1,-3){\line(0,1){6}}
}

\put(30,-2){\mbox{$:=$}}

\put(10,0){
\qbezier(50,20)(75,-40)(97,-4)  \qbezier(102,4)(105,9)(110,20)
\qbezier(50,-20)(55,-9)(58,-4) \qbezier(63,4)(85,40)(110,-20)
\put(55,9){\vector(1,-2){2}} \put(90,-13){\vector(1,1){2}} \put(109,18){\vector(1,2){2}}
\put(105,-9){\vector(-1,2){2}} \put(70,13){\vector(-1,-1){2}} \put(51,-18){\vector(-1,-2){2}}
}

\put(140,-2){\mbox{$=$}}

\put(0,0){

\put(100,65){
\put(67,-60){\vector(1,0){40}}
\put(107,-70){\vector(-1,0){40}}
\put(115,-67){\mbox{$-\ \ (q^{1-N}-q^{-N-1})$}}
\put(-25,0){
\put(233,-50){\vector(0,-1){30}}
\put(243,-80){\vector(0,1){30}}
}
}}}

}

\end{picture}
\caption{\footnotesize
The planar decomposition of the positive (in the first line) and negative (in the second line) lock vertices in the {\it vertical} framing.
}\label{fig:pladeco}
\end{figure}
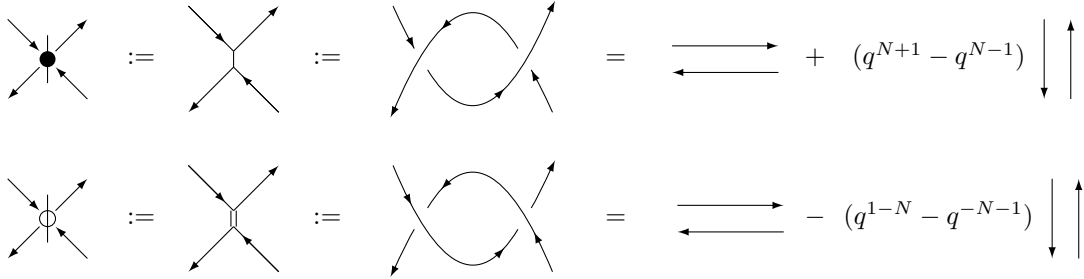

The decomposition in Fig.\,\ref{fig:pladeco-3-hyp} can be generalized to an arbitrary $N$ for a special class of bipartite links fully glued from lock vertices, see Fig.\,\ref{fig:pladeco}. In the HOMFLY case, it is crucial that two vertices are connected antiparallely. Bipartite calculus of \cite{ALM} represents the fundamental HOMFLY--PT polynomial as a sum
\begin{equation}\label{bipexp}
H^{\cal L}(A,q) = {\rm Fr}\cdot \sum_{a,b,c}  D^a \phi^b \bar\phi^c = A^{-2(n_\bullet - n_\circ)} \sum_k \sum_{i=0}^{n_\bullet} \sum_{j=0}^{n_\circ} {\cal N}_{ijk} D_N^k \phi^i \bar \phi^j
\end{equation}
where
\be
\phi = q^{N+1}-q^{N-1}, \ \ \ \ \bar\phi = -q^{1-N}+q^{-1-N} , \ \ \ \  D_N =\frac{ q^N-q^{-N} }{q-q^{-1}}\,.
\label{bippars}
\ee
The sum is over planar $2^{n_\bullet+ n_\circ}$ resolutions of $n_\bullet$ positive and $n_\circ$ negative antiparallel locks, see Fig.\,\ref{fig:pladeco}, and in the first equality all items come with unit coefficients.
However, some terms can coincide, and  if we  sum over different non-negative triples $i$, $j$, $k$,
then the coefficients are non-negative integers ${\cal N}_{ijk}$.
%Non-negativity of powers $i,j,k$ and of the coefficients ${\cal N}_{ijk}$ are the two positivity properties of PD.

What we do now, we substitute explicit expressions for $\phi$ and $\bar\phi$ as in Fig.\,\ref{fig:pladeco}\,, i.e. split two summands $\;\horr + \phi \;)\,(\;$ into three ones $\;\horr - q^{N-1}\;)\,( \, +\, q^{N+1}\;)\,(\;$ (and analogously for the mirror lock):
\begin{equation}\label{H-3-hyp}
    H^{\cal L}(A,q) = A^{-2(n_\bullet - n_\circ)} \sum_k \sum_{i=0}^{n_\bullet} \sum_{j=0}^{n_\circ} \sum_{n=0}^i \sum_{m=0}^j \begin{pmatrix}
        i \\ n
    \end{pmatrix}\begin{pmatrix}
        j \\ m
    \end{pmatrix}{\cal N}_{ijk} D_N^k \left(q^{N+1}\right)^{n-m}\left(-q^{N+1}\right)^{i-n-j+m}
\end{equation}
where $\begin{pmatrix}
        i \\ n
    \end{pmatrix} = \frac{i!}{n!(i-n)!}$ are binomial coefficients. Thus, we obtain the sum over vertices  of the peculiar 3-hypercube with $3^{n_\bullet + n_\circ}$ vertices, and in particular, pictures Fig.\,\ref{fig:Hopf-J-complex},\,\ref{fig:tw-trefoil-J-complex} are also generalized to an arbitrary $N$ in this way.

\subsection{Khovanov polynomial}

The Khovanov polynomial is the categorification ($T$-deformation) of the Jones polynomial. The method for its computation consists of the following steps~\cite{dolotin2013introduction,2506.08721}.

\paragraph{Step 1: 2-hypercube.} Draw the $2$-hypercube consisting of $2^n$ vertices and $2^{n-1}n$ edges, as described in Section~\ref{sec:Jones}.

\paragraph{Step 2: spaces.} The formula for the Jones polynomial can be rewritten in the form 
\begin{equation}\label{J-2-hyp-qdim}
    J^{\cal L}(q) = (-)^{n_\circ} q^{n_\bullet - 2 n_\circ} \sum_{r}^{2^n} (-q)^{h(r)} \dim_q(V^{\smallotimes \nu(r)})\,, \quad \dim_q(V) = D_2 = q + q^{-1}\,.
\end{equation}
So that one associates $q$-graded 2-dimensional vector space $V = \langle \vth_1,\,\vth_2 \rangle$ to each closed cycle in a resolution. The basis vectors have the gradings ${\rm grad}(\vth_1)=q^{+1}$ and ${\rm grad}(\vth_2)=q^{-1}$. It is also convenient to label these spaces by the numbers of edges of the corresponding cycles.

\paragraph{Step 3: differentials.} To each arrow of the hypercube, we associate the following morphisms:
\begin{equation}\label{Kh-morphisms}
\begin{aligned}
    \Delta &= \vth_2^b \vth_2^c \frac{\partial}{\partial \vth_2^a} + (\vth_2^b \vth_1^c + \vth_1^b \vth_2^c)\frac{\partial}{\partial \vth_1^a}\;:\quad V_a \; \mapsto \; V_b \otimes V_c\,, \\
    m &= \vth_2^a \left( \frac{\partial^2}{\partial \vth_1^b \partial \vth_2^c} + \frac{\partial^2}{\partial \vth_2^b \partial \vth_1^c} \right) + \vth_1^a \frac{\partial^2}{\partial \vth_1^b \partial \vth_1^c}\;:\quad V_b \otimes V_c \; \mapsto \; V_a\,.
\end{aligned}
\end{equation}
They are enumerated by $[\alpha_1\dots \alpha_{m-1}\,\star\, \alpha_{m+1}\dots \alpha_n]$ where the star stands at exactly the place where the mapped space label changes by unity. In order to provide the differentials property $\hat{d}_{i+1}\hat{d}_{i}=0$, with each edge, one also associates the sign factor
\begin{equation}\label{sign-morph}
    {\rm sign} = (-1)^{\alpha_1 + \dots + \alpha_{m-1}}\,.
\end{equation}
Differentials $\hat{d}_i$ are sums of operators~\eqref{Kh-morphisms} with appropriate signs~\eqref{sign-morph} acting between direct sums of spaces sharing the same fixed height $h$.

\paragraph{Step 4: kernels and images.} Then, we calculate kernels, images, cohomologies of differentials and their quantum dimensions. 

\paragraph{Step 5: Khovanov polynomial.} The Khovanov polynomial is expressed through the quantum dimensions of cohomologies ${\cal H}_i = {\rm Ker}(\hat{d}_{i})\slash {\rm Im}(\hat{d}_{i-1})$ as follows: 
\begin{equation}\label{Kh-def}
    {\rm Kh}^{\cal L}(q,T) = q^{n_\bullet}\cdot (T q^2)^{-n_\circ} \sum_{i=0}^n (qT)^i \dim_q \cH_i^{\cal L} = q^{n_\bullet}\cdot (T q^2)^{-n_\circ} \sum_{i=0}^n (qT)^i \left( \dim_q\Ker(\hat{d}_{i}^{\cal L}) - \dim_q\Im(\hat{d}_{i-1}^{\cal L}) \right)\,.
\end{equation}
The Euler characteristic can be alternatively rewritten in terms of cohomologies, so that we return to the Jones polynomial at the particular point $T=-1\,$: $J^{\cal L}(q)={\rm Kh}^{\cal L}(q,T=-1)$.

\setcounter{equation}{0}
\section{Reduction of Khovanov $(N=2)$ complexes}\label{sec:KhRed}

In this section, we provide examples of reduction of Khovanov complexes of $2^{2n}$ vertices to complexes of $3^n$ vertices where $n$ is the number of doubled crossings. These examples are presented in our Grassmann-operator formalism~\cite{dolotin2013introduction} and are consistent with our general result for the Khovanov--Rozansky polynomials~\cite{2506.08721}. We deal with the more conventional (but less convenient for us) matrix differentials~\cite{bar2002khovanov,bar2005khovanov,bar2007fast} in Appendix~\ref{sec:matrix-form}.

\subsection{One bipartite vertex}\label{sec:oper-1-bip-vert}

\noindent Let us write the maps between resolutions of the 2-unknot made from one bipartite vertex, see Fig.\,\ref{fig:bip-vert-links}. 
The conventional Kauffman--Khovanov (KK) complex is
\begin{equation}\label{2unknotcomplex}
    0 \quad \longrightarrow \quad [\bigcirc] \quad \overset{\hat{d}_0^{\,\text{2-unknot}}}{\xrightarrow{\hspace*{1cm}}} \quad 2[\bigcirc\bigcirc] \quad \overset{\hat{d}_1^{\,\text{2-unknot}}}{\xrightarrow{\hspace*{1cm}}} \quad [\bigcirc\bigcirc\bigcirc] \quad \longrightarrow \quad 0
\end{equation}
For the Hopf link, the KK complex is
\begin{equation}
\label{Hopfcomplex}
    0 \quad \longrightarrow \quad [\bigcirc \bigcirc] \quad \overset{\hat{d}_0^{\,\text{Hopf}}}{\xrightarrow{\hspace*{0.8cm}}} \quad 2[\bigcirc] \quad \overset{\hat{d}_1^{\,\text{Hopf}}}{\xrightarrow{\hspace*{0.8cm}}} \quad [\bigcirc \bigcirc] \quad \longrightarrow \quad 0
\end{equation}
see also the hypercube in Fig.\,\ref{fig:Hopf-hypercube}.

\subsubsection{Hopf link}

Let us rewrite the complex~\eqref{Hopfcomplex} in the Grassmann basis: 
\begin{equation}
    0 \quad \longrightarrow \quad \langle \eta_2,\, \theta_2 \rangle \otimes \langle \eta'_2,\, \theta'_2 \rangle \quad \overset{\hat{d}_0^{\,\text{Hopf}}}{\longrightarrow} \quad  \langle \eta_4,\, \theta_4 \rangle \oplus \langle \eta'_4,\, \theta'_4 \rangle \quad \overset{\hat{d}_1^{\,\text{Hopf}}}{\longrightarrow} \quad \langle \bar\eta_2,\, \bar\theta_2 \rangle \otimes \langle \bar\eta'_2,\, \bar\theta'_2 \rangle \quad \longrightarrow 0
\end{equation}
The differentials are 
\begin{equation}\label{dif-Hopf}
\begin{aligned}
    \hat{d}_0^{\,\text{Hopf}} = \underbrace{\left( \theta_4 + \theta'_4 \right)}_{\theta_4^+}\left( \frac{\partial}{\partial \eta_2} \frac{\partial}{\partial \theta'_2} + \frac{\partial}{\partial \theta_2} \frac{\partial}{\partial \eta'_2} \right) + \underbrace{\left( \eta_4 + \eta'_4 \right)}_{\eta_4^+} \frac{\partial}{\partial \eta_2} \frac{\partial}{\partial \eta'_2}&= \theta_4^+ \left( \frac{\partial}{\partial \eta_2} \frac{\partial}{\partial \theta'_2} + \frac{\partial}{\partial \theta_2} \frac{\partial}{\partial \eta'_2} \right) + \eta_4^+ \frac{\partial}{\partial \eta_2} \frac{\partial}{\partial \eta'_2}\,, \\
    \hat{d}_1^{\,\text{Hopf}} = \bar\theta_2 \bar\theta'_2 \underbrace{\left( \frac{\partial}{\partial \theta_4} - \frac{\partial}{\partial \theta'_4} \right)}_{\frac{\partial}{\partial \theta_4^-}} + \left( \bar\theta_2 \bar\eta'_2 + \bar\eta_2 \bar\theta'_2 \right)\underbrace{\left( \frac{\partial}{\partial \eta_4} - \frac{\partial}{\partial \eta'_4} \right)}_{\frac{\partial}{\partial \eta_4^-}}&= \bar\theta_2 \bar\theta'_2 \frac{\partial}{\partial \theta_4^-}
    + \left( \bar\theta_2 \bar\eta'_2 + \bar\eta_2 \bar\theta'_2 \right) \frac{\partial}{\partial \eta_4^-}\,.
\end{aligned}
\end{equation}
% The reduction of $\mathfrak{d}_2^{\circ \circ}$ is performed according to the change of basis~\eqref{Hopf-diag-basis} and vanishing of the ``odd'' components. Namely, the $\theta_4$-, $\eta_4$-dependent terms in $\mathfrak{d}_2^{\circ \circ}$ turn to zero and the following substitutions are made
% \begin{equation}
% \begin{aligned}
%     \theta_2 \eta'_2 \; &\rightarrow \; -\eta_2 \theta'_2 \\ 
%     \theta_2 \; &\rightarrow \; 0
% \end{aligned}
% \end{equation}
% The result is
% \begin{equation}
%     \mathfrak{d}_2^{\circ \circ} \quad \longrightarrow \quad -\left( -\eta_2 \theta'_2 + \eta_2 \theta'_2 \right)\frac{\partial}{\partial \eta'_4}=0
% \end{equation}
% Note that in the basis $\langle \eta_4^+ = \eta_4 + \eta'_4,\,\theta_4^+ = \theta_4 + \theta'_4 \rangle \oplus \langle \eta_4^- = \eta_4 - \eta'_4,\,\theta_4^- = \theta_4 - \theta'_4 \rangle$ the differential $\mathfrak{d}_2^{\circ \circ}$ provides the isomorphism $V^- \rightarrow V^-$. Thus, $V^-=\langle \eta_4^-,\, \theta_4^- \rangle$ can be eliminated, and the complex reduces to
In the second column in~\eqref{dif-Hopf} we explicitly separated the variables $\theta_4^\pm = \theta_4 \pm \theta_4^-$ and $\eta_4^\pm = \eta_4 \pm \eta_4^-$
what makes explicit the nilpotent property $  \hat{d}_1^{\,\text{Hopf}}\hat{d}_0^{\,\text{Hopf}} = 0$.
Indeed, the zeroth operator $\hat{d}_0^{\,\text{Hopf}}$ produces only functions of $\theta_4^+$ and $\eta_4^+$,
while the first one $\hat{d}_1^{\,\text{Hopf}}$ is non-vanishing only on functions of $\theta_4^-$ and $\eta_4^-$.
This also implies explicit reduction by substitution of the space $ \quad 2[\bigcirc]$ of linear functions in four variables
$\theta_4^\pm,\eta_4^\pm$
by $ \quad [\bigcirc]$, consisting of linear functions of $\theta_4^+,\eta_4^+$ only: the reduced complex is
\begin{equation}
    \varnothing \quad \longrightarrow \quad \langle \eta_2,\, \theta_2 \rangle \otimes \langle \eta'_2,\, \theta'_2 \rangle \quad \overset{m}{\longrightarrow} \quad \langle \eta_4^+,\, \theta_4^+ \rangle \quad \overset{0}{{\color{Green}\longrightarrow}} \quad \langle \bar\eta_2^+,\, \bar\theta_2^+ \rangle \quad \longrightarrow \quad \varnothing
\end{equation}
Thus, the initial complex in the first line reduces to the complex in the second line:
\be
    \varnothing \quad \longrightarrow \quad [\bigcirc \bigcirc] \quad \overset{\hat{d}_0^{\,\text{Hopf}}}{\longrightarrow} \quad \slashed{2}[\bigcirc] \quad \overset{\hat{d}_1^{\,\text{Hopf}}}{\longrightarrow} \quad [\cancel{\bigcirc} \bigcirc] \quad \longrightarrow \quad \varnothing
    \nn
    \ee
    \be
    \downarrow  \nn
    \ee
    \be
    \varnothing \quad \longrightarrow \quad [\bigcirc \bigcirc] \quad \overset{m}{\longrightarrow} \quad [\bigcirc] \quad \overset{0}{{\color{Green}\longrightarrow}} \quad [\bigcirc] \quad \longrightarrow  \quad \varnothing
\ee
which is the simplest example of the complex of 3 vertices introduced in this paper, compare with the hypercube in Fig.\,\ref{fig:Hopf-J-complex}. Note that the reduction is possible because the vanished subcomplex is exact, i.e. does not contribute to the cohomologies.

\subsubsection{2-unknot}\label{sec:2-unknot}

Let us rewrite the complex~\eqref{2unknotcomplex} in the Grassmann basis: 
\begin{equation}\label{Gr-complex}
    0 \; \longrightarrow \; 
    \langle \eta_4,\, \theta_4 \rangle \; \overset{\hat{d}_0^{\,\text{2-unknot}}}{\longrightarrow} \; \left(\langle \eta_1,\, \theta_1 \rangle \otimes \langle \eta_3,\, \theta_3 \rangle\right)\oplus \left(\langle \eta'_3,\, \theta'_3 \rangle \otimes \langle \eta'_1,\, \theta'_1 \rangle\right) \; \overset{\hat{d}_1^{\,\text{2-unknot}}}{\longrightarrow} \;  \langle \eta_1,\, \theta_1 \rangle \otimes \langle \eta_2,\, \theta_2 \rangle \otimes \langle \eta'_1,\, \theta'_1 \rangle
    \; \longrightarrow \; 0
%\langle \eta_1 \eta_4,\, \eta_1 \theta_4,\, \theta_1 \eta_4,\, \theta_1 \theta_4 \rangle
\end{equation}
In these variables, the differentials are
\begin{equation}\label{dif-2-unknot}
\begin{aligned}
    \hat{d}_0^{\,\text{2-unknot}} &= \underbrace{(\theta_1 \theta_3 + \theta'_3\theta'_1)}_{\theta_1^+ \theta_3^+}\frac{\partial}{\partial \theta_4} + \Big(\underbrace{\theta_1 \eta_3 + \eta'_3\theta'_1}_{\theta_1^+ \eta_3^+} + \underbrace{\eta_1 \theta_3 + \theta'_3\eta'_1}_{\eta_1^+ \theta_3^+} \Big)\frac{\partial}{\partial \eta_4}\,, \\
    \hat{d}_1^{\,\text{2-unknot}} &= \theta_1 \theta_2 \frac{\partial}{\partial\theta'_3} + \left(\theta_1 \eta_2 + \eta_1 \theta_2\right)\frac{\partial}{\partial\eta'_3} - \theta'_1 \theta_2 \frac{\partial}{\partial\theta_3} - \left(\theta'_1 \eta_2 + \eta'_1 \theta_2\right)\frac{\partial}{\partial\eta_3}\,.
\end{aligned}
\end{equation}
Again, the zeroth differential $\hat{d}_0^{\,\text{2-unknot}}$ produces only functions of $\theta_1^+ \theta_3^+$, $\eta_1^+ \theta_3^+$, $\theta_1^+ \eta_3^+$, $\eta_1^+ \eta_3^+$. Thus, it is convenient to explicitly separate the corresponding differentials $\frac{\partial}{\partial\theta_1^+} \frac{\partial}{\partial\theta_3^+}$, $\frac{\partial}{\partial\eta_1^+} \frac{\partial}{\partial\theta_3^+}$, $\frac{\partial}{\partial\theta_1^+} \frac{\partial}{\partial\eta_3^+}$, $\frac{\partial}{\partial\eta_1^+} \frac{\partial}{\partial\eta_3^+}$ from their $(-)$-counterparts. The initial differential $\hat{d}_1^{\,\text{2-unknot}} =\Delta^{\prime} \otimes \mathds{1}' + \mathds{1} \otimes (-\Delta)$ in~\eqref{dif-2-unknot} is linear in derivatives, but it is easily transformed to quadratic form by the expansion of the unity operators in order to provide the subsequent reduction:
\begin{equation}
\begin{aligned}
    \mathds{1}' = \eta'_1 \frac{\partial}{\partial \eta'_1} + \theta'_1 \frac{\partial}{\partial \theta'_1} \quad {\rm and} \quad \mathds{1} = \eta_1 \frac{\partial}{\partial \eta_1} + \theta_1 \frac{\partial}{\partial \theta_1}\,.
\end{aligned}
\end{equation}
% However, in order to provide the Gaussian elimination as in Section~\ref{sec:matrix-form}, one should distinguish all the spaces: 
% \begin{equation}\label{Gr-complex}
%     \varnothing \; \longrightarrow \; 
%     \langle \eta_5,\, \theta_5 \rangle \; \overset{d_1^{\circ \circ}}{\longrightarrow} \; \left(\langle \eta_1,\, \theta_1 \rangle \otimes \langle \eta_4,\, \theta_4 \rangle\right)\oplus \left(\langle \eta'_4,\, \theta'_4 \rangle \otimes \langle \eta'_1,\, \theta'_1 \rangle\right) \; \overset{d_2^{\circ \circ}}{\longrightarrow} \;  \langle \eta_3,\, \theta_3 \rangle \otimes \langle \eta_2,\, \theta_2 \rangle \otimes \langle \eta'_3,\, \theta'_3 \rangle
%     \; \longrightarrow \; \varnothing
% \end{equation}
Then, the second differential takes the form
\begin{equation}\label{d2-ext}
\begin{aligned}
    \hat{d}_1^{\,\text{2-unknot}} &= \theta_1 \theta_2 \underbrace{\left( \eta'_1 \frac{\partial}{\partial \eta'_1} + \theta'_1 \frac{\partial}{\partial \theta'_1} \right)}_{\mathds{1}'}\frac{\partial}{\partial\theta'_3} + \left(\theta_1 \eta_2 + \eta_1 \theta_2\right)\underbrace{\left( \eta'_1 \frac{\partial}{\partial \eta'_1} + \theta'_1 \frac{\partial}{\partial \theta'_1} \right)}_{\mathds{1}'}\frac{\partial}{\partial\eta'_3} - \\
    &- \theta'_1 \theta_2 \underbrace{\left( \eta_1 \frac{\partial}{\partial \eta_1} + \theta_1 \frac{\partial}{\partial \theta_1} \right)}_{\mathds{1}} \frac{\partial}{\partial\theta_3} - \left(\theta'_1 \eta_2 + \eta'_1 \theta_2\right)\underbrace{\left( \eta_1 \frac{\partial}{\partial \eta_1} + \theta_1 \frac{\partial}{\partial \theta_1} \right)}_{\mathds{1}}\frac{\partial}{\partial\eta_3} = \\
    %&= - \left(\eta_3 \theta_2 \eta'_3 + \eta_3\eta_2 \theta'_3\right)\frac{\partial}{\partial\eta_1}\frac{\partial}{\partial\eta_4} - \eta_3 \theta_2 \theta'_3 \frac{\partial}{\partial\eta_1}\frac{\partial}{\partial\theta_4} - \left( \theta_3 \eta_2 \theta'_3 + \theta_3 \theta_2 \eta'_3 \right)\frac{\partial}{\partial\theta_1}\frac{\partial}{\partial\eta_4} - \theta_3 \theta_2 \theta'_3 \frac{\partial}{\partial\theta_1}\frac{\partial}{\partial\theta_4} + \\
    %& + \left(\theta_3 \eta_2 \eta'_3 + \eta_3 \theta_2 \eta'_3 \right)\frac{\partial}{\partial\eta'_4}\frac{\partial}{\partial\eta'_1} + \left( \theta_3 \eta_2 \theta'_3 + \eta_3 \theta_2 \theta'_3 \right)\frac{\partial}{\partial\eta'_4}\frac{\partial}{\partial\theta'_1} + \theta_3 \theta_2 \eta'_3 \frac{\partial}{\partial\theta'_4}\frac{\partial}{\partial\eta'_1} + \theta_3 \theta_2 \theta'_3 \frac{\partial}{\partial\theta'_4}\frac{\partial}{\partial\theta'_1}= \\
    &= \eta_1 \theta_2 \eta'_1 \underbrace{\left( \frac{\partial}{\partial \eta'_1} \frac{\partial}{\partial \eta'_3} - \frac{\partial}{\partial \eta_3} \frac{\partial}{\partial \eta_1}\right)}_{\frac{\partial}{\partial \eta_1^-} \frac{\partial}{\partial \eta_3^-}} + \eta_1 \theta_2 \theta'_1 \underbrace{\left( \frac{\partial}{\partial \eta'_1} \frac{\partial}{\partial \theta'_3} - \frac{\partial}{\partial \eta_3} \frac{\partial}{\partial \theta_1}\right)}_{\frac{\partial}{\partial \eta_1^-} \frac{\partial}{\partial \theta_3^-}} + \theta_1 \theta_2 \eta'_1 \underbrace{\left( \frac{\partial}{\partial \theta'_1} \frac{\partial}{\partial \eta'_3} - \frac{\partial}{\partial \theta_3} \frac{\partial}{\partial \eta_1}\right)}_{\frac{\partial}{\partial \theta_1^-} \frac{\partial}{\partial \eta_3^-}} + \\
    &+ \theta_1 \theta_2 \theta'_1 \underbrace{\left( \frac{\partial}{\partial \theta'_1} \frac{\partial}{\partial \theta'_3} - \frac{\partial}{\partial \theta_3} \frac{\partial}{\partial \theta_1}\right)}_{\frac{\partial}{\partial \theta_1^-} \frac{\partial}{\partial \theta_3^-}} + \theta_1 \eta_2 \eta'_1 \frac{\partial}{\partial \eta'_1} \frac{\partial}{\partial \eta'_3} + \theta_1 \eta_2 \theta'_1 \frac{\partial}{\partial \theta'_1} \frac{\partial}{\partial \eta'_3} - \eta_1 \eta_2 \theta'_1 \frac{\partial}{\partial \eta_3} \frac{\partial}{\partial \eta_1} - \theta_1 \eta_2 \theta'_1 \frac{\partial}{\partial \eta_3} \frac{\partial}{\partial \theta_1} = \\
    &= \left( \eta_1 \theta_2 \eta'_1 + \frac{1}{2} \theta_1 \eta_2 \eta'_1 + \frac{1}{2} \eta_1 \eta_2 \theta'_1 \right)\frac{\partial}{\partial \eta_1^-} \frac{\partial}{\partial \eta_4^-} + \left( \eta_1 \theta_2 \theta'_1 + \frac{1}{2} \theta_1 \eta_2 \theta'_1 \right)\frac{\partial}{\partial \eta_1^-} \frac{\partial}{\partial \theta_3^-} + \\ 
    &+ \left( \theta_1 \theta_2 \eta'_1 + \frac{1}{2} \theta_1 \eta_2 \theta'_1 \right)\frac{\partial}{\partial \theta_1^-} \frac{\partial}{\partial \eta_3^-} + \theta_1 \theta_2 \theta'_1 \frac{\partial}{\partial \theta_1^-} \frac{\partial}{\partial \theta_3^-} + \\
    &+ \boxed{\frac{1}{2}\left(\theta_1 {\color{blue} \eta_2} \eta'_1 - \eta_1 {\color{blue} \eta_2} \theta'_1 \right) \frac{\partial}{\partial \eta_1^+} \frac{\partial}{\partial \eta_3^+} + \frac{1}{2}\theta_1 {\color{blue} \eta_2} \theta'_1 \frac{\partial}{\partial \theta_1^+} \frac{\partial}{\partial \eta_3^+} - \frac{1}{2}\theta_1 {\color{blue} \eta_2} \theta'_1 \frac{\partial}{\partial \eta_1^+} \frac{\partial}{\partial \theta_3^+}\,.}
\end{aligned}
\end{equation}
To transfer to the last equality, we have made the following decompositions:
\begin{equation}
\begin{aligned}
    \frac{\partial}{\partial \eta'_1} \frac{\partial}{\partial \eta'_3} &= \frac{1}{2}\frac{\partial}{\partial \eta_1^+} \frac{\partial}{\partial \eta_3^+} + \frac{1}{2}\frac{\partial}{\partial \eta_1^-} \frac{\partial}{\partial \eta_3^-}\,, \quad \frac{\partial}{\partial \eta_3} \frac{\partial}{\partial \eta_1} = \frac{1}{2}\frac{\partial}{\partial \eta_1^+} \frac{\partial}{\partial \eta_3^+} - \frac{1}{2}\frac{\partial}{\partial \eta_1^-} \frac{\partial}{\partial \eta_3^-}\,, \\
    \frac{\partial}{\partial \theta'_1} \frac{\partial}{\partial \eta'_3} &= \frac{1}{2}\frac{\partial}{\partial \theta_1^+} \frac{\partial}{\partial \eta_3^+} + \frac{1}{2}\frac{\partial}{\partial \theta_1^-} \frac{\partial}{\partial \eta_3^-}\,, \quad \frac{\partial}{\partial \eta_3} \frac{\partial}{\partial \theta_1} = \frac{1}{2} \frac{\partial}{\partial \eta_1^+} \frac{\partial}{\partial \theta_3^+} - \frac{1}{2} \frac{\partial}{\partial \eta_1^-} \frac{\partial}{\partial \theta_3^-}\,.
\end{aligned}
\end{equation}
The differential $\hat{d}_0^{\,\text{2-unknot}}$ maps only to the $(+)$-subspace. This implies explicit reduction by the substitution of the space $ \quad 2[\bigcirc \bigcirc]$ of linear functions in eight variables
$\theta_1^\pm\theta_3^\pm$, $\eta_1^\pm\eta_3^\pm$, $\eta_1^\pm\theta_3^\pm$, $\theta_1^\pm\eta_3^\pm$ 
by $ \quad [\bigcirc\bigcirc]$ consisting of linear functions of $\theta_1^+\theta_3^+$, $\eta_1^+\eta_3^+$, $\eta_1^+\theta_3^+$, $\theta_1^+\eta_3^+$ only. Also note that the differential $\hat{d}_1^{\,\text{2-unknot}}$ leaves only $\eta_2$-dependent elements, and the image is four dimensional. Thus, $\eta_2$ can be just omitted, and the reduced complex is
  
%and $V \otimes \theta_2 \otimes V$:
% \begin{equation}
%     d_2^{\circ \circ} \Big( \eta_4^- \eta_1^-,\, \eta_4^- \theta_1^-,\, \theta_4^- \eta_1^-  ,\, \theta_4^- \theta_1^-  \Big) = \Big( \underline{\eta_3 \theta_2 \eta'_3 + \frac{1}{2} \eta_3 \eta_2 \theta'_3 + \frac{1}{2} \theta_3 \eta_2 \eta'_3,\, \underline{\eta_3 \theta_2 \theta'_3} + \theta_3 \eta_2 \theta'_3,\, \underline{\theta_3 \theta_2 \eta'_3},\, \underline{\theta_3 \theta_2 \theta'_3} \Big)
% \end{equation}
%and the complex~\eqref{Gr-complex} can be reduced to
\begin{equation}\label{red-2-unknot-complex}
    \varnothing \quad \longrightarrow \quad \langle \eta_4,\, \theta_4 \rangle \quad \overset{\Delta}{\longrightarrow} \quad \langle \eta_1^+,\, \theta_1^+ \rangle \otimes \langle \eta_3^+,\, \theta_3^+ \rangle \quad \overset{\rm Sh}{{\color{blue}\longrightarrow}} \quad \langle \eta_1,\, \theta_1 \rangle \otimes \langle \eta'_1,\, \theta'_1 \rangle
\end{equation}
with
\begin{equation}\label{Sh-morph}
\begin{aligned}
    \Delta &= \theta_1^+ \theta_3^+ \frac{\partial}{\partial \theta_4} + \left( \theta_1^+ \eta_3^+ + \eta_1^+ \theta_3^+ \right) \frac{\partial}{\partial \eta_4}\,, \\
    m &= \theta_2 \left( \frac{\partial}{\partial\eta_1} \frac{\partial}{\partial\theta'_1} + \frac{\partial}{\partial\theta_1} \frac{\partial}{\partial\eta'_1} \right) + \eta_2 \frac{\partial}{\partial\eta_1} \frac{\partial}{\partial\eta'_1}\,, \\
    {\rm Sh} &= \theta_1 \theta'_1 \left(\frac{\partial}{\partial \theta_1^+} \frac{\partial}{\partial \eta_3^+} - \frac{\partial}{\partial \eta_1^+} \frac{\partial}{\partial \theta_3^+}\right) + \left(\theta_1 \eta'_1 - \eta_1 \theta'_1 \right) \frac{\partial}{\partial \eta_1^+} \frac{\partial}{\partial \eta_3^+}\,.
\end{aligned}
\end{equation}
Note that the new Sh morphism is the same as in~\cite{2506.08721} for $N=2$. The check of the differential properties ${\rm Sh}\,\Delta = 0$ and $m\,{\rm Sh} = 0$ is straightforward. 

The full reduction procedure in schematic notations is:
\be
    \varnothing \quad \longrightarrow \quad [\bigcirc] \quad \overset{\hat{d}_0^{\,\text{2-unknot}}}{\longrightarrow} \quad \slashed{2}[\bigcirc\bigcirc] \quad \overset{\hat{d}_1^{\,\text{2-unknot}}}{\longrightarrow} \quad [\bigcirc\cancel{\bigcirc}\bigcirc] \quad \longrightarrow \quad \varnothing
    \nn
    \ee
    \be
    \downarrow  \nn
    \ee
    \be
    \varnothing \quad \longrightarrow \quad [\bigcirc] \quad \overset{\Delta}{\longrightarrow} \quad [\bigcirc\bigcirc] \quad \overset{\rm Sh}{{\color{blue}\longrightarrow}} \quad  [\bigcirc\bigcirc] \quad \longrightarrow \quad \varnothing
\ee
Here again, the deleted subcomplex is exact, and from the 4-hypercube, we have obtained the 3-hypercube. Note that the reduction of the hypercubes is the same as in Section~\ref{sec:2-unknot-red}.

\subsection{Trefoil knot}

In this subsection, we provide the reductions of the trefoil complexes. The twist trefoil knot from Fig.\,\ref{fig:bip-vert-links} consists only from double vertices. Thus, the $2^4$-hypercube can be reduced to $3^2$-hypercube. Due to the antiparallel connection of vertices, this reduction also works beyond the considered $N=2$ case~\cite{2506.08721}. We also consider the example of the torus (2-strand trefoil), see Fig.\,\ref{fig:trefoil-hypercube}. It is glued from one single crossing and one double crossing. Still, this double crossing allows for the reduction: $2^3$-hypercube can be simplified to $(2\cdot 3)$-hypercube.

\subsubsection{Bipartite (twist) version}\label{sec:twist-trefoil}

The hypercube is present in Fig.\,\ref{fig:4-trefoil-hypercube}. The complex can be written in short as
\begin{equation}\label{trefoil-complex}
    \varnothing \; \longrightarrow \;[\bigcirc]\; \overset{\hat{d}_0}{\longrightarrow} \; 4[\bigcirc\bigcirc]\; \overset{\hat{d}_1}{\longrightarrow} \; 2[\bigcirc\bigcirc\bigcirc] \oplus 4[\bigcirc]\; \overset{\hat{d}_2}{\longrightarrow} \; 4[\bigcirc\bigcirc]\; \overset{\hat{d}_3}{\longrightarrow} \; [\bigcirc\bigcirc\bigcirc] \; \longrightarrow \; \varnothing
\end{equation}
or in explicit Grassmann variables:
{\tiny \begin{equation}
    \begin{tabular}{c|c|c|c|c|c|c}
        & & & $ \langle \eta_3^{(2)},\,\theta_3^{(2)} \rangle \otimes \langle \eta_2,\,\theta_2 \rangle \otimes \langle \eta_3^{(1)},\,\theta_3^{(1)} \rangle $   \\  
        & & & $\oplus$ \\
          & $ \langle \eta_{5}^{(1)},\,\theta_{5}^{(1)} \rangle \otimes \langle \eta_{3}^{(1)},\,\theta_{3}^{(1)} \rangle $ & & $ \langle \eta_8^{(1)},\,\theta_8^{(1)} \rangle $ & & $ \langle \eta_{6}^{(1)},\,\theta_{6}^{(1)} \rangle \otimes \langle \eta_{2},\,\theta_{2} \rangle $  \\ 
         & $\oplus$ & & $\oplus$ & & $\oplus$ \\
         & $ \langle \eta_{3}^{(2)},\,\theta_{3}^{(2)} \rangle \otimes \langle \eta_{5}^{(2)},\,\theta_{5}^{(2)} \rangle $ & & $ \langle \eta_8^{(2)},\,\theta_8 \rangle $ &  & $ \langle \eta_{6}^{(2)},\,\theta_{6}^{(2)} \rangle \otimes \langle \eta_{2},\,\theta_{2} \rangle $ \\
         $\langle \eta_8,\, \theta_8\rangle \overset{\hat{d}_0}{\rightarrow}$ & $\oplus$ & $\overset{\hat{d}_1}{\rightarrow}$ & $\oplus$ & $\overset{\hat{d}_2}{\rightarrow}$ & $\oplus$ & $\overset{\hat{d}_3}{\rightarrow}\langle \eta_{4},\,\theta_{4} \rangle \otimes \langle \eta'_{2},\,\theta'_{2} \rangle \otimes \langle \eta_{2},\,\theta_{2} \rangle $ \\
         & $ \langle \eta_{5}^{(3)},\,\theta_{5}^{(3)} \rangle \otimes \langle \eta_{3}^{(3)},\,\theta_{3}^{(3)} \rangle $ & & $ \langle \eta_8^{(3)},\,\theta_8^{(3)} \rangle $ &  & $ \langle \eta'_{2},\,\theta'_{2} \rangle \otimes \langle \eta_{6}^{(3)},\,\theta_{6}^{(3)} \rangle $ \\
         & $\oplus$ & & $\oplus$ & & $\oplus$ \\
         & $ \langle \eta_{3}^{(4)},\,\theta_{3}^{(4)} \rangle \otimes \langle \eta_{5}^{(4)},\,\theta_{5}^{(4)} \rangle $ & & $ \langle \eta_{8}^{(4)},\,\theta_{8}^{(4)} \rangle $ &  & $ \langle \eta'_{2},\,\theta'_{2} \rangle \otimes \langle \eta_{6}^{(4)},\,\theta_{6}^{(4)} \rangle $ \\
          & & & $\oplus$ \\
          & & & $ \langle \eta_{3}^{(4)},\,\theta_{3}^{(4)} \rangle \otimes \langle \eta'_{2},\,\theta'_{2} \rangle \otimes \langle \eta_{3}^{(3)},\,\theta_{3}^{(3)} \rangle $ 
    \end{tabular}
\end{equation}}
The zeroth differential $\hat{d}_0 = \Delta^{(1)} + \Delta^{(2)} + \Delta^{(3)} + \Delta^{(4)}$ is
\begin{equation}
\begin{aligned}
    \hat{d}_0 &= 
    %\sum\limits_{j=2}^5 \left[ \theta_j \theta'_j \frac{\partial}{\partial \theta_1} + \left( \theta_j \eta'_j + \eta_j \theta'_j \right) \frac{\partial}{\partial \eta_1} \right] = 
    \left( \theta_{3}^{\scaleto{(1)+(2)}{5pt}} \theta^{\scaleto{(1)+(2)}{5pt}}_{5} + \theta_{3}^{\scaleto{(3)+(4)}{5pt}} \theta^{\scaleto{(3)+(4)}{5pt}}_{5} \right) \frac{\partial}{\partial \theta_8} + \left( \eta_{3}^{\scaleto{(1)+(2)}{5pt}} \theta^{\scaleto{(1)+(2)}{5pt}}_{5} + \theta_{3}^{\scaleto{(1)+(2)}{5pt}} \eta^{\scaleto{(1)+(2)}{5pt}}_{5} + \eta_{3}^{\scaleto{(3)+(4)}{5pt}} \theta^{\scaleto{(3)+(4)}{5pt}}_{5} + \theta_{3}^{\scaleto{(3)+(4)}{5pt}} \eta^{\scaleto{(3)+(4)}{5pt}}_{5} \right) \frac{\partial}{\partial \eta_8} = \\
    &= \boxed{\Delta^{\scaleto{(1)+(2)}{5pt}} + \Delta^{\scaleto{(3)+(4)}{5pt}}: \quad V_8 \; \mapsto \; (V_3^{\scaleto{(1)+(2)}{5pt}} \otimes V_5^{\scaleto{(1)+(2)}{5pt}}) \oplus (V_3^{\scaleto{(3)+(4)}{5pt}} \otimes V_5^{\scaleto{(3)+(4)}{5pt}})\,.}
\end{aligned}
\end{equation}
where we split the variables:
\begin{equation}
\begin{aligned}
    \theta_{3}^{\scaleto{(j)\pm(j+1)}{5.5pt}} \theta^{\scaleto{(j)\pm(j+1)}{5.5pt}}_{5} &= \theta_{3}^{(j)} \theta^{(j)}_{5} \pm \theta_{5}^{(j+1)} \theta^{(j+1)}_{3}\,, \quad \eta_{3}^{\scaleto{(j)\pm(j+1)}{5.5pt}} \eta^{\scaleto{(j)\pm(j+1)}{5.5pt}}_{5} = \eta_{3}^{(j)} \eta^{(j)}_{5} \pm \eta_{5}^{(j+1)} \eta^{(j+1)}_{3}\,, \\
    \theta_{3}^{\scaleto{(j)\pm(j+1)}{5.5pt}} \eta^{\scaleto{(j)\pm(j+1)}{5.5pt}}_{5} &= \theta_{3}^{(j)} \eta^{(j)}_{5} \pm \theta_{5}^{(j+1)} \eta^{(j+1)}_{3}\,, \quad \eta_{3}^{\scaleto{(j)\pm(j+1)}{5.5pt}} \theta^{\scaleto{(j)\pm(j+1)}{5.5pt}}_{5} = \eta_{3}^{(j)} \theta^{(j)}_{5} \pm \eta_{5}^{(j+1)} \theta^{(j+1)}_{3}\,.
\end{aligned}
\end{equation}
The operator $\hat{d}_0$ maps only to eight-dimensional space $2(V^+ \otimes V^+)$ of functions of the $(+)$-variables. Thus, we can get rid of the spaces $\left(\langle \eta_{3}^{\scaleto{(1)-(2)}{5pt}},\, \theta_{3}^{\scaleto{(1)-(2)}{5pt}} \rangle \otimes \langle \eta_{5}^{\scaleto{(1)-(2)}{5pt}},\, \theta_{5}^{\scaleto{(1)-(2)}{5pt}} \rangle\right) \oplus \left(\langle \eta_{3}^{\scaleto{(3)-(4)}{5pt}},\, \theta_{3}^{\scaleto{(3)-(4)}{5pt}} \rangle \otimes \langle \eta_{5}^{\scaleto{(3)-(4)}{5pt}},\, \theta_{5}^{\scaleto{(3)-(4)}{5pt}} \rangle\right) $, and in the differential $\hat{d}_1$, we take into account only the $(+)$-differentials. In the suboperator $\Delta^{(1)} \otimes \mathds{1}^{(1)} + \mathds{1}^{(2)} \otimes (-\Delta^{(2)}) + \Delta^{(3)} \otimes \mathds{1}^{(3)} + \mathds{1}^{(4)} \otimes (-\Delta^{(4)})$, we again expand the unities $\mathds{1}^{(j)}= \theta_3^{(j)}\frac{\partial}{\partial \theta_3^{(j)}} + \eta_3^{(j)}\frac{\partial}{\partial \eta_3^{(j)}}$ to further reduce the complex. The resulting differential is

{\footnotesize \begin{equation}
\begin{aligned}
    \hat{d}_1 &= \frac{1}{2}\sum\limits_{i=1}^4 \eta_8^{(i)} \left( \frac{\partial}{\partial \eta_{3}^{\scaleto{(1)+(2)}{5pt}}}\frac{\partial}{\partial \eta_{5}^{\scaleto{(1)+(2)}{5pt}}} - \frac{\partial}{\partial \eta_{3}^{\scaleto{(3)+(4)}{5pt}}}\frac{\partial}{\partial \eta_{5}^{\scaleto{(3)+(4)}{5pt}}} \right) + \\ 
    &+ \frac{1}{2}\sum\limits_{i=1}^4 \theta_8^{(i)} \left( \frac{\partial}{\partial \eta_{3}^{\scaleto{(1)+(2)}{5pt}}}\frac{\partial}{\partial \theta_{5}^{\scaleto{(1)+(2)}{5pt}}} + \frac{\partial}{\partial \theta_{3}^{\scaleto{(1)+(2)}{5pt}}}\frac{\partial}{\partial \eta_{5}^{\scaleto{(1)+(2)}{5pt}}} - \frac{\partial}{\partial \eta_{3}^{\scaleto{(3)+(4)}{5pt}}}\frac{\partial}{\partial \theta_{5}^{\scaleto{(3)+(4)}{5pt}}} - \frac{\partial}{\partial \theta_{3}^{\scaleto{(3)+(4)}{5pt}}}\frac{\partial}{\partial \eta_{5}^{\scaleto{(3)+(4)}{5pt}}} \right) + \\
    &+ \frac{1}{2}\theta_3^{(2)} {\color{blue} \eta_2} \theta_3^{(1)} \left( \frac{\partial}{\partial \theta_{3}^{\scaleto{(1)+(2)}{5pt}}}\frac{\partial}{\partial \eta_{5}^{\scaleto{(1)+(2)}{5pt}}} - \frac{\partial}{\partial \eta_{3}^{\scaleto{(1)+(2)}{5pt}}}\frac{\partial}{\partial \theta_{5}^{\scaleto{(1)+(2)}{5pt}}} \right) + \frac{1}{2}\left(\theta_3^{(2)} {\color{blue} \eta_2} \eta_3^{(1)} - \eta_3^{(2)} {\color{blue} \eta_2} \theta_3^{(1)} \right) \frac{\partial}{\partial \eta_{3}^{\scaleto{(1)+(2)}{5pt}}}\frac{\partial}{\partial \eta_{5}^{\scaleto{(1)+(2)}{5pt}}} + \\
    &+ \frac{1}{2}\theta_3^{(4)} {\color{blue} \eta_2} \theta_3^{(3)} \left( \frac{\partial}{\partial \theta_{3}^{\scaleto{(3)+(4)}{5pt}}}\frac{\partial}{\partial \eta_{5}^{\scaleto{(3)+(4)}{5pt}}} - \frac{\partial}{\partial \eta_{3}^{\scaleto{(3)+(4)}{5pt}}}\frac{\partial}{\partial \theta_{5}^{\scaleto{(3)+(4)}{5pt}}} \right) + \frac{1}{2}\left(\theta_3^{(4)} {\color{blue} \eta_2} \eta_3^{(3)} - \eta_3^{(4)} {\color{blue} \eta_2} \theta_3^{(3)} \right) \frac{\partial}{\partial \eta_{3}^{\scaleto{(3)+(4)}{5pt}}}\frac{\partial}{\partial \eta_{5}^{\scaleto{(3)+(4)}{5pt}}} \cong \\
    &\cong \boxed{m^{\scaleto{(1)+(2)}{5pt}} -m^{\scaleto{(3)+(4)}{5pt}} + {\rm Sh}^{\scaleto{(1)+(2)}{5pt}} + {\rm Sh}^{\scaleto{(3)+(4)}{5pt}}: \; (V_3^{\scaleto{(1)+(2)}{5pt}} \otimes V_5^{\scaleto{(1)+(2)}{5pt}}) \oplus (V_3^{\scaleto{(3)+(4)}{5pt}} \otimes V_5^{\scaleto{(3)+(4)}{5pt}}) \; \mapsto \; (V_3^{(2)} \otimes V_3^{(1)}) \oplus V_8^{\scaleto{(1)+(2)+(3)+(4)}{5pt}}\oplus (V_3^{(4)} \otimes V_3^{(3)})\,.}
\end{aligned}
\end{equation}}

\begin{figure}[h!]
\centering
\begin{picture}(100,110)(40,-5)

\put(-150,50){\mbox{\bf Kh}}

\put(-87,50){\mbox{$0\;\overset{0}{\longrightarrow} \; [\overset{4}{\bigcirc}]_{[00]}^{\scaleto{\{0\}}{5.5pt}}$}}
\put(-35,65){\vector(1,1){20}}
\put(-35,40){\vector(1,-1){20}}
\put(-40,75){\mbox{$\Delta$}}
\put(-25,65){\mbox{\scriptsize $[0\star]$}}
\put(-40,25){\mbox{$\Delta$}}
\put(-25,35){\mbox{\scriptsize $[\star 0]$}}

\put(-10,90){\mbox{$[\overset{2}{\bigcirc}\overset{2'}{\bigcirc}]_{[01]}^{\scaleto{\{1\}}{5.5pt}}$}}
\put(-10,10){\mbox{$[\overset{\bar 2}{\bigcirc}\overset{\bar 2'}{\bigcirc}]_{[10]}^{\scaleto{\{1\}}{5.5pt}}$}}
\put(-3,50){\mbox{$\bigoplus$}}

\put(10,0){

\put(60,90){\mbox{$[\overset{{\color{blue} 2}}{\bigcirc}\overset{{\color{blue} 2'}}{\bigcirc}]_{[02]}^{\scaleto{\{3\}}{5.5pt}}$}}
\put(60,10){\mbox{$[\overset{{\color{blue} \bar 2}}{\bigcirc}\overset{{\color{blue} \bar 2'}}{\bigcirc}]_{[20]}^{\scaleto{\{3\}}{5.5pt}}$}}
{\color{blue} \put(31,93){\vector(1,0){25}}
\put(31,13){\vector(1,0){25}}}
\put(35,100){\mbox{\small Sh}}
\put(45,17){\mbox{\scriptsize $[\star 0]$}}
\put(45,84){\mbox{\scriptsize $[0\star]$}}
\put(35,0){\mbox{\small Sh}}

\put(68,32){\mbox{$\bigoplus$}}
\put(68,72){\mbox{$\bigoplus$}}
\put(63,50){\mbox{$[\overset{4'}{\bigcirc}]_{[11]}^{\scaleto{\{2\}}{5.5pt}}$}}
\put(32,83){\vector(1,-1){27}}
\put(30,60){\mbox{$m$}}
\put(51,69){\mbox{\scriptsize $[\star 1]$}}
\put(32,20){\vector(1,1){27}}
\put(25,35){\mbox{$-m$}}
\put(51,32){\mbox{\scriptsize $[1\star]$}}

\put(10,0){

\put(110,0){\mbox{$m$}}
\put(110,100){\mbox{$m$}}
\put(101,93){\vector(1,0){25}}
\put(101,13){\vector(1,0){25}}
\put(130,90){\mbox{$[\overset{{\color{Green} 4}}{\bigcirc}]_{[12]}^{\scaleto{\{4\}}{5.5pt}}$}}
\put(130,10){\mbox{$[\overset{{\color{Green} 4'}}{\bigcirc}]_{[21]}^{\scaleto{\{4\}}{5.5pt}}$}}
\put(133,50){\mbox{$\bigoplus$}}
{\color{Green} 
\put(100,57){\vector(1,1){27}}
\put(100,50){\vector(1,-1){27}}}
\put(98,17){\mbox{\scriptsize $[2\star]$}}
\put(98,84){\mbox{\scriptsize $[\star 2]$}}
\put(120,35){\mbox{$0$}}
\put(120,60){\mbox{$0$}}
\put(98,32){\mbox{\scriptsize $[\star 1]$}}
\put(96,69){\mbox{\scriptsize $[1 \star]$}}

\put(10,0){

{\color{Green} 
\put(150,83){\vector(1,-1){27}}
\put(150,22){\vector(1,1){27}}}
\put(182,50){\mbox{$[\bigcirc]_{[22]}^{\scaleto{\{6\}}{5.5pt}}\;\overset{0}{\longrightarrow} \; 0$}}

\put(165,25){\mbox{$0$}}
\put(165,75){\mbox{$0$}}
\put(145,65){\mbox{\scriptsize $[\star 2]$}}
\put(145,35){\mbox{\scriptsize $[2 \star]$}}
}
}
}
    
\end{picture}
\caption{\footnotesize The complex for the trefoil knot in the bipartite presentation. We color in blue the arrows for the morphisms Sh and the corresponding labellings of spaces. The Sh morphisms go after the $\Delta$ morphisms, while the zero morphisms go after the $m$ morphisms. The arrows and labels in green correspond to zero morphisms. Spaces are enumerated by $[\alpha_1\, \alpha_2]$ with $\alpha_1=0,\,1,\,2$ corresponding to smoothings shown in Fig.\,\ref{fig:pladeco-3-hyp}. Arrows are labelled by $[\alpha_1 \star]$ or $[\star \alpha_2]$ depending on the change of the enumerators of spaces.}
\label{fig:tw-trefoil-complex}
\end{figure}
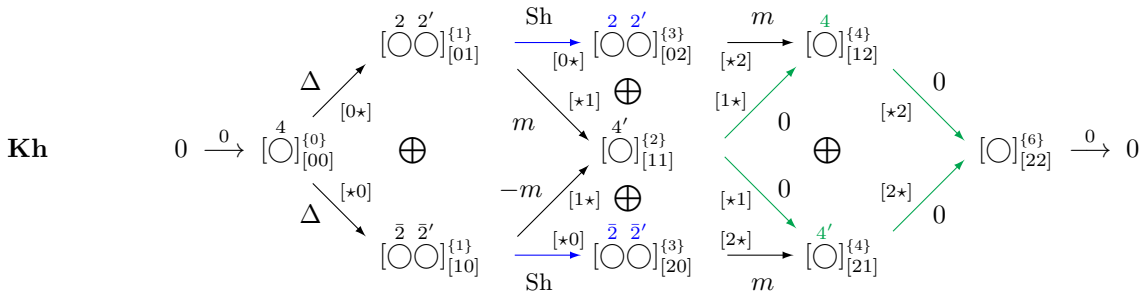

\noindent One can easily check that the vectors $\sum\limits_{i=1}^4 \eta_8^{(i)}$, $\sum\limits_{i=1}^4 \theta_8^{(i)}$ lie in the kernel of the unreduced $\hat{d}_2$. %Moreover, the differential $d_2$ gives only $\eta'_6$ and $\eta'_{11}$ dependent vectors, and $\theta'_6$, $\theta'_{11}$ dependent ones can be eliminated. 
Thus, one can consider only the remaining part of the operator $\hat{d}_2$: 
\begin{equation}
\begin{aligned}
    \hat{d}_2 &= \theta_6^{\scaleto{(1)+(2)}{5pt}}\left( \frac{\partial}{\partial \eta_3^{(2)}} \frac{\partial}{\partial \theta_3^{(1)}} + \frac{\partial}{\partial \theta_3^{(2)}} \frac{\partial}{\partial \eta_3^{(1)}} \right) + \eta_6^{\scaleto{(1)+(2)}{5pt}} \frac{\partial}{\partial \eta_3^{(2)}} \frac{\partial}{\partial \eta_3^{(1)}} + \\
    &+\theta_6^{\scaleto{(3)+(4)}{5pt}}\left( \frac{\partial}{\partial \eta_3^{(4)}} \frac{\partial}{\partial \theta_3^{(3)}} + \frac{\partial}{\partial \theta_3^{(4)}} \frac{\partial}{\partial \eta_3^{(3)}} \right) + \eta_6^{\scaleto{(3)+(4)}{5pt}} \frac{\partial}{\partial \eta_3^{(4)}} \frac{\partial}{\partial \eta_3^{(3)}} = \\
    &= \boxed{m^{\scaleto{(1)+(2)}{5pt}} + m^{\scaleto{(3)+(4)}{5pt}} + 0^{\scaleto{(1)+(2)}{5pt}} + 0^{\scaleto{(3)+(4)}{5pt}}: \quad (V_3^{(2)} \otimes V_3^{(1)}) \oplus V_8^{\scaleto{(1)+(2)+(3)+(4)}{5pt}}\oplus (V_3^{(4)} \otimes V_3^{(3)}) \; \mapsto \; V_6^{\scaleto{(1)+(2)}{5pt}} \oplus V_6^{\scaleto{(3)+(4)}{5pt}}\,.}
\end{aligned}
\end{equation}
Here, we again introduce the variables: 
\begin{equation}
    \eta_6^{\scaleto{(j)\pm(j+1)}{5.5pt}} = \eta_6^{(j)} \pm \eta_6^{(j+1)}\,, \quad \theta_6^{\scaleto{(j)\pm(j+1)}{5.5pt}} = \theta_6^{(j)} \pm \theta_6^{(j+1)}\,. 
\end{equation}
%??? Note that $d_3$ effectively maps to 2-dimensional spaces, because the $d_2$ operator leaves us only the peculiar combination of vectors:
%{\scriptsize \begin{equation}
 %   d_2\Big( \theta_6 \eta'_6 \theta''_6,\, - \eta_6 \eta'_6 \theta''_6 + \theta_6 \eta'_6 \eta''_6,\, \theta_{11} \eta'_{11} \theta''_{11},\, - \eta_{11} \eta'_{11} \theta''_{11} + \theta_{11} \eta'_{11} \eta''_{11} \Big)= \Big( \theta_{12,13}^+ \theta_{12,13}^{\prime +},\, \theta_{12,13}^+ \eta_{12,13}^{\prime +}-\eta_{12,13}^+ \theta_{12,13}^{\prime +},\, \theta_{14,15}^+ \theta_{14,15}^{\prime +},\, \theta_{14,15}^+ \eta_{14,15}^{\prime +} - \eta_{14,15}^+ \theta_{14,15}^{\prime +}\Big)
%\end{equation}}
The remaining differential depends only on $(-)$-variables:
\begin{equation}
\begin{aligned}
    \hat{d}_3 &= \theta_4 \theta'_2 \frac{\partial}{\partial \theta_6^{\scaleto{(1)-(2)}{5pt}}} + \left( \theta_4 \eta'_2 + \eta_4 \theta'_2 \right) \frac{\partial}{\partial \eta_6^{\scaleto{(1)-(2)}{5pt}}} + \theta_4 \theta_2 \frac{\partial}{\partial \theta_6^{\scaleto{(3)-(4)}{5pt}}} + \left( \theta_4 \eta_2 + \eta_4 \theta_2 \right) \frac{\partial}{\partial \eta_6^{\scaleto{(3)-(4)}{5pt}}}\,,
\end{aligned}
\end{equation}
and therefore, acts as zero on the invariant $(+)$-subspaces. Thus, the resulting reduced complex is:
\begin{equation}
\begin{tabular}{c|c|c|c|c|c|c}
    & & & $\langle \eta_3^{(2)},\, \theta_3^{(2)} \rangle \otimes \langle \eta_3^{(1)},\, \theta_3^{(1)} \rangle $  \\
    & $\langle \eta_{3}^{\scaleto{(1)+(2)}{5pt}},\, \theta_{3}^{\scaleto{(1)+(2)}{5pt}} \rangle \otimes \langle \eta_{5}^{\scaleto{(1)+(2)}{5pt}},\, \theta_{5}^{\scaleto{(1)+(2)}{5pt}} \rangle$ & & $\oplus$ & & $\langle \eta_{6}^{\scaleto{(1)+(2)}{5pt}},\, \theta_{6}^{\scaleto{(1)+(2)}{5pt}} \rangle$ \\
   $\varnothing \rightarrow \langle \eta_8,\, \theta_8 \rangle \rightarrow$ & $\oplus$ & $\rightarrow$ & $\langle \eta_{8}^{\scaleto{(1)+(2)+(3)+(4)}{5pt}},\, \theta_{8}^{\scaleto{(1)+(2)+(3)+(4)}{5pt}} \rangle$ & $\rightarrow$ & $\oplus$ & $\rightarrow \varnothing$ \\
    & $\langle \eta_{3}^{\scaleto{(3)+(4)}{5pt}},\, \theta_{3}^{\scaleto{(3)+(4)}{5pt}} \rangle \otimes \langle \eta_{5}^{\scaleto{(3)+(4)}{5pt}},\, \theta_{5}^{\scaleto{(3)+(4)}{5pt}} \rangle$ & & $\oplus$ & & $\langle \eta_{6}^{\scaleto{(3)+(4)}{5pt}},\, \theta_{6}^{\scaleto{(3)+(4)}{5pt}} \rangle$ \\
    & & & $\langle \eta_3^{(4)},\, \theta_3^{(4)} \rangle \otimes \langle \eta_3^{(3)},\, \theta_3^{(3)} \rangle $ 
\end{tabular}
\end{equation}
The full reduction procedure in schematic notations is:
\be
   \varnothing \; \longrightarrow \;[\bigcirc]\; \overset{\hat{d}_0}{\longrightarrow} \; (2+\slashed{2})[\bigcirc\bigcirc]\; \overset{\hat{d}_1}{\longrightarrow} \; 2[\bigcirc\cancel{\bigcirc}\bigcirc] \oplus \slashed{4}[\bigcirc]\; \overset{\hat{d}_2}{\longrightarrow} \; (2+\slashed{2})[\cancel{\bigcirc}\bigcirc]\; \overset{\hat{d}_3}{\longrightarrow} \; [\cancel{\bigcirc\bigcirc}\bigcirc] \; \longrightarrow \; \varnothing
    \nn
    \ee
    \be
    \downarrow  \nn
    \ee
    \be\label{trefoil-red-complex}
    \varnothing \; \longrightarrow \; [\bigcirc] \; \overset{\Delta+\Delta}{\longrightarrow} \; 2[\bigcirc\bigcirc] \; \overset{{\rm Sh} + m - m + {\rm Sh}}{\longrightarrow} \;  [\bigcirc\bigcirc]\oplus[\bigcirc]\oplus[\bigcirc\bigcirc] \; \overset{m+0+0+m}{\longrightarrow} \; 2[\bigcirc] \; \overset{0}{\longrightarrow} \; [\bigcirc] \; \longrightarrow \; \varnothing
\ee
One can get sure that the discarded subcomplex is indeed exact. In the full form, the resulting complex is shown in Fig.\,\ref{fig:tw-trefoil-complex}. The structure of the 3-hypercube is better seen in Fig.\,\ref{fig:tw-trefoil-J-complex}.

\subsubsection{2-strand (torus) representation}\label{sec:2-str-trefoil}

The hypercube is shown in Fig.\,\ref{fig:trefoil-hypercube}. The complex is
\begin{equation}
\begin{tabular}{c|c|c|c|c}
    & $\langle \eta_6,\, \theta_6 \rangle$ & & $\langle \eta''_4,\, \theta''_4 \rangle \otimes \langle \eta''_2,\, \theta''_2 \rangle$ \\ 
    & $\oplus$ & & $\oplus$ \\
    $\varnothing \rightarrow \langle \eta_3,\, \theta_3 \rangle\otimes \langle \eta'_3,\, \theta'_3 \rangle \overset{\hat{d}_0}{\rightarrow}$ & $\langle \eta'_6,\, \theta'_6 \rangle$ & $\overset{\hat{d}_1}{\rightarrow}$ & $\langle \eta_4,\, \theta_4 \rangle \otimes \langle \eta_2,\, \theta_2 \rangle$ & $\overset{\hat{d}_2}{\rightarrow} \langle \eta_2,\, \theta_2 \rangle \otimes \langle \eta'_2,\, \theta'_2 \rangle \otimes \langle \eta''_2,\, \theta''_2 \rangle\rightarrow \varnothing$ \\
    & $\oplus$ & & $\oplus$ \\
    & $\langle \eta''_6,\, \theta''_6 \rangle$ & & $\langle \eta'_2,\, \theta'_2 \rangle \otimes \langle \eta'_4,\, \theta'_4 \rangle$
\end{tabular}
\end{equation}
where the differentials are
\begin{equation}
\begin{aligned}
    \hat{d}_0 &= (\theta_6 + \theta'_6 + \theta''_6)\left( \frac{\partial}{\partial \eta_3}\frac{\partial}{\partial \theta'_3} + \frac{\partial}{\partial \theta_3}\frac{\partial}{\partial \eta'_3} \right) + ( \eta_6 + \eta'_6 + \eta''_6 )\frac{\partial}{\partial \eta_3}\frac{\partial}{\partial \eta'_3}\,, \\
    \hat{d}_1 &= ( \theta'_4 \theta'_2 - \theta''_4 \theta''_2 )\frac{\partial}{\partial \theta_6} + ( \eta'_4 \theta'_2 + \theta'_4 \eta'_2 - \theta''_4 \eta''_2 - \eta''_4 \theta''_2 )\frac{\partial}{\partial \eta_6} + \\
    &+ ( \theta''_4 \theta''_2 - \theta_4 \theta_2 )\frac{\partial}{\partial \theta'_6} + ( \eta''_4 \theta''_2 + \theta''_4 \eta''_2 - \theta_4 \eta_2 - \eta_4 \theta_2 )\frac{\partial}{\partial \eta'_6} + \\
    &+ ( \theta_4 \theta_2 - \theta'_4 \theta'_2 )\frac{\partial}{\partial \theta''_6} + ( \eta_4 \theta_2 + \theta_4 \eta_2 - \theta'_4 \eta'_2 - \eta'_4 \theta'_2 )\frac{\partial}{\partial \eta''_6}\,, \\
    \hat{d}_2 &= \theta_2 \theta'_2 \frac{\partial}{\partial \theta''_4} + (\eta_2 \theta'_2 + \theta_2 \eta'_2)\frac{\partial}{\partial \eta''_4} + \theta''_2 \theta_2 \frac{\partial}{\partial \theta'_4} + (\eta''_2 \theta_2 + \theta''_2 \eta_2)\frac{\partial}{\partial \eta'_4} + \theta'_2 \theta''_2 \frac{\partial}{\partial \theta_4} + (\eta'_2 \theta''_2 + \theta'_2 \eta''_2)\frac{\partial}{\partial \eta_4}\,.
\end{aligned}
\end{equation}
Kernels and images are
{\small \begin{equation}
\begin{aligned}
    {\rm Ker}(\hat{d}_0) &= \{ \theta_3 \theta'_3, (\theta_3 \eta'_3 - \eta_3 \theta'_3) \}\,,\quad {\rm Im}(\hat{d}_0)=\{ \theta_6 + \theta'_6 + \theta''_6,\, \eta_6 + \eta'_6 + \eta''_6 \} \,, \quad \dim_q({\cal H}_0) = \dim_q{\rm Ker}(\hat{d}_0)= q^{-2}+1\,, \\
    {\rm Ker}(\hat{d}_1) &= {\rm Im}(\hat{d}_0)\,, \quad {\rm Im}(\hat{d}_1) = \{ \theta'_4 \theta'_2 - \theta''_4 \theta''_2,\, \theta''_4 \theta''_2 - \theta_4 \theta_2,\, \eta'_4 \theta'_2 + \theta'_4 \eta'_2 - \theta''_4 \eta''_2 - \eta''_4 \theta''_2,\, \eta''_4 \theta''_2 + \theta''_4 \eta''_2 - \theta_4 \eta_2 - \eta_4 \theta_2 \}\,, \\ 
    {\rm Ker}(\hat{d}_2) &= \{ \theta_4\theta_2 - \theta'_4\theta'_2,\, \theta_4\theta_2 - \theta''_4\theta''_2,\, \eta'_4 \theta'_2 - \theta_4 \eta_2 - \theta''_4 \eta''_2,\, \eta_4 \theta_2 - \theta'_4 \eta'_2 - \theta''_4\eta''_2,\, \eta''_4 \theta''_2 - \theta_4 \eta_2 - \theta'_4 \eta'_2 \}\,, \; {\rm CoIm}(\hat{d}_2) = \{ \eta_2 \eta'_2 \eta''_2 \}\,.
\end{aligned}
\end{equation}}
Note that the differential $d_2$ maps to the eight-dimensional space. Let us choose its basis as:
\begin{equation}
    \underbrace{\langle \eta_4^- \eta_2^-,\, \eta_4^- \theta_2^-,\, \theta_4^- \eta_2^-,\, \theta_4^- \theta_2^- \rangle}_{V^-} \oplus \underbrace{\langle \eta'_4,\, \theta'_4 \rangle  \eta'_2}_{V'_4} \oplus \underbrace{\langle \eta_4^{\prime -}\theta_2^{\prime -} + \theta_4^{\prime -}\eta_2^{\prime -},\, \theta_4^{\prime -}\theta_2^{\prime -} \rangle}_{V^{\prime -}}\,,
\end{equation}
and also change the basis in $V\oplus V\oplus V$ as follows:
\begin{equation}
     \underbrace{\langle \eta_6,\, \theta_6}_{V_6} \rangle \oplus \underbrace{\langle \eta'_6,\, \theta'_6 \rangle}_{V'_6} \oplus \underbrace{\langle \eta_6^+,\, \theta_6^+ \rangle}_{V_6^+} 
\end{equation}
where
\begin{equation}
\begin{aligned}
    \eta_6^+ &= \eta_6 + \eta'_6 + \eta''_6\,, \quad \theta_6^+ = \theta_6 + \theta'_6 + \theta''_6
    \eta_4^- \eta_2^- &= \eta_4 \eta_2 - \eta''_4 \eta''_2\,, \quad \theta_4^- \theta_2^- = \theta_4 \theta_2 - \theta''_4 \theta''_2\,, \\
    \eta_4^- \theta_2^- &= \eta_4 \theta_2 - \theta''_4 \eta''_2\,, \quad \theta_4^- \eta_2^- = \theta_4 \eta_2 - \eta''_4 \theta''_2\,, \\
    \eta_4^{\prime -} \eta_2^{\prime -} &= \eta'_4 \eta'_2 - \eta''_4 \eta''_2\,, \quad \theta_4^{\prime -} \theta_2^{\prime -} = \theta'_4 \theta'_2 - \theta''_4 \theta''_2\,, \\
    \eta_4^{\prime -} \theta_2^{\prime -} &= \eta'_4 \theta'_2 - \theta''_4 \eta''_2\,, \quad \theta_4^{\prime -} \eta_2^{\prime -} = \theta'_4 \eta'_2 - \eta''_4 \theta''_2\,.
\end{aligned}
\end{equation}
In this basis
\begin{equation}
\begin{aligned}
    \hat{d}_0 &= \theta_6^+ \left( \frac{\partial}{\partial \eta_3}\frac{\partial}{\partial \theta'_3} + \frac{\partial}{\partial \theta_3}\frac{\partial}{\partial \eta'_3} \right) + \eta_6^+\frac{\partial}{\partial \eta_3}\frac{\partial}{\partial \eta'_3}=\boxed{0 + 0' + m^{+}:\quad V_3 \otimes V'_3 \; \mapsto \; V_6 \oplus V'_6 \oplus V_6^+\,,} \\
    \hat{d}_1 &= \theta_4^{\prime -} \theta_2^{\prime -} \frac{\partial}{\partial \theta_6} + \left( \eta_4^{\prime -}\theta_2^{\prime -} + \theta_4^{\prime -}\eta_2^{\prime -} \right) \frac{\partial}{\partial \eta_6} - \theta_4^- \theta_2^- \frac{\partial}{\partial \theta'_6} - \left( \eta_4^{-}\theta_2^{-} + \theta_4^{-}\eta_2^{-} \right) \frac{\partial}{\partial \eta'_6} + 0 \cdot \frac{\partial}{\partial \theta_6^+} + 0 \cdot \frac{\partial}{\partial \eta_6^+}= \\
    &= \boxed{\mathds{1}^{} + (-\Delta^{\prime}) + 0^+: \quad V_6 \oplus V'_6 \oplus V_6^+ \; \mapsto \; V^{\prime -} \oplus V^- \oplus V'_4\,,} \\
    \hat{d}_2 &= \left( \eta_2 {\color{blue} \eta'_2} \theta''_2 + \theta_2 {\color{blue} \eta'_2} \eta''_2 \right) \frac{\partial}{\partial \eta_2^{-}} \frac{\partial}{\partial \eta_4^{-}} + \theta_2 {\color{blue} \eta'_2} \theta''_2 \left(\frac{\partial}{\partial \theta_2^{-}} \frac{\partial}{\partial \eta_4^{-}} - \frac{\partial}{\partial \eta_2^{-}} \frac{\partial}{\partial \theta_4^{-}}\right) + \left( \theta_2 \eta''_2 + \eta_2 \theta''_2 \right) \frac{\partial}{\partial \eta'_4} + \theta_2 \theta''_2 \frac{\partial}{\partial \theta'_4} = \\
    &= \boxed{0^{\prime -} + {\color{blue} \eta'_2} {\rm Sh} + \Delta: \quad V^{\prime -} \oplus V^- \oplus V'_4 \; \mapsto \; {\color{blue} \eta'_2} \langle \eta_2\eta''_2,\, \eta_2\theta''_2,\, \theta_2\eta''_2,\, \theta_2\theta''_2 \rangle\,.}
\end{aligned}
\end{equation}
Here in $\hat{d}_2$ we have left only the operators acting on $V^{\prime -} \oplus V^- \oplus V'_4$ and expanded the unity operators. Note that the resulting complex contains the subcomplex
\begin{equation}
    V_3 \otimes V'_3 \; \overset{0}{\mapsto} \; V_6 \; \overset{\mathds{1}}{\mapsto} \; V^{\prime -} \; \overset{0^{\prime -}}{\mapsto} \; {\color{blue} \eta'_2} \langle \eta_2\eta''_2,\, \eta_2\theta''_2,\, \theta_2\eta''_2,\, \theta_2\theta''_2 \rangle
\end{equation}
which does not contribute to the cohomologies, and thus, can be excluded. Therefore, the resulting reduced complex is
\begin{equation}
\begin{tabular}{ccccccc}
     & $\overset{0}{{\color{Green}\longrightarrow}}$  & $\langle \eta'_6,\, \theta'_6 \rangle $ & $\overset{-\Delta}{\longrightarrow}$ & $\langle \eta_4^-,\, \theta_4^- \rangle \otimes \langle \eta_2^-,\, \theta_2^- \rangle$ & $\overset{\rm Sh}{{\color{blue}\longrightarrow}}$ \\
     $\langle \eta_3,\, \theta_3 \rangle \otimes \langle \eta'_3,\, \theta'_3 \rangle$ & & $\oplus $ & & $\oplus$ & & $\langle \eta_2,\, \theta_2 \rangle \otimes \langle \eta''_2,\, \theta''_2 \rangle$ \\
     & $\overset{m}{\longrightarrow}$ & $\langle \eta_6^+,\, \theta_6^+ \rangle$ & $\overset{0}{{\color{Green}\longrightarrow}}$ & $\langle \eta'_4,\, \theta'_4 \rangle$ & $\overset{\Delta}{\longrightarrow}$
\end{tabular}
\end{equation}
Note that in the HOMFLY (an arbitrary $N$) case, the initial complex cannot be reduced in such a way, because the orientation of strands in a torus knot is parallel.

\setcounter{equation}{0}
\section{Conclusion}\label{sec:concl}

In this paper, we have worked with knots, including double vertices. In the formalism of odd differential operators, we have provided examples of reductions of Khovanov complexes for such knots. In the case of bipartite links fully constructed from lock tangles being double vertices with antiparallel orientation, our result reproduces the expectations coming from~\cite{2506.08721,2508.05191,lanina2026khovanov,galakhov2026reductions}. 

Further research can focus on reductions of more difficult tangles, allowing computations of the Khovanov polynomial to be faster. Another interesting question is whether there are such simplifications of complexes for colored Khovanov polynomials.

\section*{Acknowledgments}

We are grateful for enlightening discussions to R. Stepanov. 

This work was funded by the RSF grant No.24-12-00178.

\appendix

\setcounter{equation}{0}
\section{Matrix form reductions}\label{sec:matrix-form}

In this appendix, we provide the analogous reduction procedure in the matrix form~\cite{bar2002khovanov}. Actually, the generic algorithm is present in~\cite{bar2007fast}, but here we make straightforward explicit calculations. 

In the matrix formalism, each circle is equipped with the 2-dimensional vector space $V=\langle v_{+},\,v_{-}\rangle$. Matrices for the co-product and multiplication operators~\eqref{Kh-morphisms}, respectively, look as follows:
\begin{equation}\label{prim-diff}
    \Delta = \begin{pmatrix}
        0 & 0 \\
        1 & 0 \\
        1 & 0 \\
        0 & 1
    \end{pmatrix},\qquad m = \begin{pmatrix}
        1 & 0 & 0 & 0 \\
        0 & 1 & 1 & 0
    \end{pmatrix}\,.
\end{equation}
These matrices are written in the bases $\langle v_+,\, v_- \rangle$ and $\langle v_+ \otimes v_+, \, v_+ \otimes v_-, \, v_- \otimes v_+, \, v_- \otimes v_- \rangle$.

\subsection{Hopf link} 

We again consider the complex~\eqref{Hopfcomplex}. The differential $\hat{d}_0^{\,\text{Hopf}} =m + m$ in the matrix form consists of two identical $4\times 2$ matrices $m$ from~\eqref{prim-diff}:
\begin{equation}\label{d0-Hopf}
    \hat{d}_0^{\,\text{Hopf}} = \begin{pmatrix}
        1 & 0 & 0 & 0 \\
        0 & 1 & 1 & 0 \\
        \hline
        1 & 0 & 0 & 0 \\
        0 & 1 & 1 & 0 
    \end{pmatrix}.
\end{equation}
The differential $\hat{d}_1^{\,\text{Hopf}} =-\Delta + \Delta$ in the matrix form: 
%it consists of two identical $2\times 4$ matrices $d_1^\bullet$ from~\eqref{prim-diff}:
\begin{equation}
    \hat{d}_1^{\,\text{Hopf}} = \left(\begin{array}{cc|cc}
        0 & 0 & 0 & 0 \\
        -1 & 0 & 1 & 0 \\
        \hline
        -1 & 0 & 1 & 0 \\
        0 & -1 & 0 & 1
    \end{array}\right).
\end{equation}
Its diagonal form in the bases
\begin{equation}\label{Hopf-diag-basis}
    \langle v_+^{(1)} + v_+^{(2)},\, v_-^{(1)} + v_-^{(2)},\, v_+^{(2)},\, v_-^{(2)} \rangle \quad {\rm and} \quad \langle v_+ \otimes v_+,\, v_{+} \otimes v_- - v_- \otimes v_+,\, v_- \otimes v_+,\, v_- \otimes v_- \rangle
\end{equation}
is
\begin{equation}\label{d1-Hopf}
    \hat{d}_1^{\,\text{Hopf}} = \left(\begin{array}{cc|cc}
        0 & 0 & 0 & 0 \\
        0 & 0 & 0 & 0 \\
        \hline
        0 & 0 & 1 & 0 \\
        0 & 0 & 0 & 1
    \end{array}\right).
\end{equation}
One can also change the basis of the remaining vector space to transform the matrix~\eqref{d0-Hopf}
\begin{equation}\label{d0-Hopf-2}
    \hat{d}_0^{\,\text{Hopf}} = \begin{pmatrix}
        1 & 0 & 0 & 0 \\
        0 & 1 & 1 & 0 \\
        \hline
        0 & 0 & 0 & 0 \\
        0 & 0 & 0 & 0 
    \end{pmatrix}.
\end{equation}
The $2\times 2$ identity matrix at the bottom of~\eqref{d1-Hopf} and the zero map in~\eqref{d0-Hopf-2} allows us to provide the elimination of exact subcomplex leading to the reduced complex:
\begin{equation}\label{Hopf-red-complex}
    \varnothing \quad \longrightarrow \quad [\bigcirc \bigcirc] \quad \overset{m}{\longrightarrow} \quad [\bigcirc] \quad \overset{0}{{\color{Green}\longrightarrow}} \quad [\bigcirc] \quad \longrightarrow \quad \varnothing
\end{equation}

\subsection{2-unknot} 

Let us now deal with the complex~\eqref{2unknotcomplex}. The differential $\hat{d}_0^{\,\text{2-unknot}} =\Delta + \Delta$ in the matrix form consists of two identical $2\times 4$ matrices $\Delta$ from~\eqref{prim-diff}:
\begin{equation}
    \hat{d}_0^{\,\text{2-unknot}} = \begin{pmatrix}
        0 & 0 \\
        1 & 0 \\
        1 & 0 \\
        0 & 1 \\
        \hline
        0 & 0 \\
        1 & 0 \\
        1 & 0 \\
        0 & 1
    \end{pmatrix}.
\end{equation}
The differential $\hat{d}_1^{\,\text{2-unknot}} =(-\mathds{1}\otimes \Delta) + (\Delta \otimes \mathds{1})$ in the matrix form is
\begin{equation}
    \hat{d}_1^{\,\text{2-unknot}} = \begin{pmatrix}
        0 & 0 & 0 & 0 & 0 & 0 & 0 & 0 \\
        -1 & 0 & 0 & 0 & 0 & 0 & 0 & 0 \\
        -1 & 0 & 0 & 0 & 1 & 0 & 0 & 0 \\
        0 & -1 & 0 & 0 & 0 & 1 & 0 & 0 \\
        0 & 0 & 0 & 0 & 1 & 0 & 0 & 0 \\
        0 & 0 & -1 & 0 & 0 & 1 & 0 & 0 \\
        0 & 0 & -1 & 0 & 0 & 0 & 1 & 0 \\
        0 & 0 & 0 & -1 & 0 & 0 & 0 & 1
    \end{pmatrix}.
\end{equation}
Note that this matrix is written in the bases $\langle v_{+}^{(1)}\otimes v_{+}^{(1)},\,v_{+}^{(1)}\otimes v_{-}^{(1)},\,v_{-}^{(1)}\otimes v_{+}^{(1)},\,v_{-}^{(1)}\otimes v_{-}^{(1)},\,v_{+}^{(2)}\otimes v_{+}^{(2)},\,v_{+}^{(2)}\otimes v_{-}^{(2)},\,v_{-}^{(2)}\otimes v_{+}^{(2)},\,v_{-}^{(2)}\otimes v_{-}^{(2)}\rangle$ and $\langle v_{+} \otimes v_{+} \otimes v_{+},\, v_{+} \otimes v_{+} \otimes v_{-},\, v_{+} \otimes v_{-} \otimes v_{+},\, v_{+} \otimes v_{-} \otimes v_{-},\, v_{-} \otimes v_{+} \otimes v_{+},\, v_{-} \otimes v_{+} \otimes v_{-},\, v_{-} \otimes v_{-} \otimes v_{+},\, v_{-} \otimes v_{-} \otimes v_{-}\rangle$. 

We claim now that this complex has a natural reduction, reflecting the bipartite nature of the underlying knot diagram.
In order to %build an algorithm, extendable to tangles, and to 
introduce the new operator $\rm Sh$ for the reduced complex, 
we must eliminate one circle in $[\bigcirc\bigcirc\bigcirc]$.
To better visualize this,
it is convenient to change the second set of basis vectors to $\langle v_{+} \otimes v_{+} \otimes v_{+},\, v_{+} \otimes v_{+} \otimes v_{-},\, v_{-} \otimes v_{+} \otimes v_{+},\, v_{-} \otimes v_{+} \otimes v_{-},\, v_{+} \otimes v_{-} \otimes v_{+},\, v_{+} \otimes v_{-} \otimes v_{-},\, v_{-} \otimes v_{-} \otimes v_{+},\, v_{-} \otimes v_{-} \otimes v_{-}\rangle$.
i.e. permute two pairs of entries from {\footnotesize $(+++,++-,\underline{+-+,+--,\underline{-++,-+-}},--+,---)$} to 
{\footnotesize $(+++,++-,\underline{\underline{-++,-+-},+-+,+--},--+,---)$}.
In this basis
\begin{equation}\label{d2-el-form}
    \hat{d}_1^{\,\text{2-unknot}} = \left(\begin{array}{cccc|cccc}
        0 & 0 & 0 & 0 & 0 & 0 & 0 & 0 \\
        -1 & 0 & 0 & 0 & 0 & 0 & 0 & 0 \\
        0 & 0 & 0 & 0 & 1 & 0 & 0 & 0 \\
        0 & 0 & -1 & 0 & 0 & 1 & 0 & 0 \\
        \hline
        -1 & 0 & 0 & 0 & 1 & 0 & 0 & 0 \\
        0 & -1 & 0 & 0 & 0 & 1 & 0 & 0 \\
        0 & 0 & -1 & 0 & 0 & 0 & 1 & 0 \\
        0 & 0 & 0 & -1 & 0 & 0 & 0 & 1
    \end{array}\right).
\end{equation}
Then, we block-diagonalize the matrix~\eqref{d2-el-form}. Namely, in the bases 
\begin{equation}
\begin{aligned}
    &\langle v_{+}^{(1)}\otimes v_{+}^{(1)}+v_{+}^{(2)}\otimes v_{+}^{(2)},\,v_{+}^{(1)}\otimes v_{-}^{(1)}+v_{+}^{(2)}\otimes v_{-}^{(2)},\,v_{-}^{(1)}\otimes v_{+}^{(1)}+v_{-}^{(2)}\otimes v_{+}^{(2)},\,v_{-}^{(1)}\otimes v_{-}^{(1)}+v_{-}^{(2)}\otimes v_{-}^{(2)},\\ 
    &v_{+}^{(2)}\otimes v_{+}^{(2)},\,v_{+}^{(2)}\otimes v_{-}^{(2)},\,v_{-}^{(2)}\otimes v_{+}^{(2)},\,v_{-}^{(2)}\otimes v_{-}^{(2)}\rangle\,, \\ 
    &\langle v_{+} \otimes v_{+} \otimes v_{+},\, v_{+} \otimes v_{+} \otimes v_{-},\, v_{-} \otimes v_{+} \otimes v_{+} - v_{+} \otimes v_{-} \otimes v_{+},\, v_{-} \otimes v_{+} \otimes v_{-} - v_{+} \otimes v_{-} \otimes v_{-},\\ 
    &v_{+} \otimes v_{-} \otimes v_{+},\, v_{+} \otimes v_{-} \otimes v_{-},\, v_{-} \otimes v_{-} \otimes v_{+},\, v_{-} \otimes v_{-} \otimes v_{-}\rangle 
\end{aligned}
\end{equation}
the matrix of $\hat{d}_1^{\,\text{2-unknot}}$ takes the form

\begin{equation}\label{d1-unknot}
    \hat{d}_1^{\,\text{2-unknot}} = \left(\begin{array}{cccc|cccc}
        0 & 0 & 0 & 0 & 0 & 0 & 0 & 0 \\
        -1 & 0 & 0 & 0 & 0 & 0 & 0 & 0 \\
        1 & 0 & 0 & 0 & 0 & 0 & 0 & 0 \\
        0 & 1 & -1 & 0 & 0 & 0 & 0 & 0 \\
        \hline
        0 & 0 & 0 & 0 & 1 & 0 & 0 & 0 \\
        0 & 0 & 0 & 0 & 0 & 1 & 0 & 0 \\
        0 & 0 & 0 & 0 & 0 & 0 & 1 & 0 \\
        0 & 0 & 0 & 0 & 0 & 0 & 0 & 1
    \end{array}\right).
\end{equation}
Changing the basis of the remaining vector space, one gets:
\begin{equation}\label{d0-unknot}
    \hat{d}_0^{\,\text{2-unknot}} = \begin{pmatrix}
        0 & 0 \\
        1 & 0 \\
        1 & 0 \\
        0 & 1 \\
        \hline
        0 & 0 \\
        0 & 0 \\
        0 & 0 \\
        0 & 0
    \end{pmatrix}.
\end{equation}
There is the identity isomorphism at the right bottom corner of the matrix~\eqref{d1-unknot} and the zero morphism at the bottom of~\eqref{d0-unknot}, thus, the corresponding subcomplex does not contribute to the cohomology. 
This means that the complex (\ref{2unknotcomplex}) can be reduced to
\begin{equation}
    \varnothing \quad \longrightarrow \quad [\bigcirc] \quad \overset{\Delta}{\longrightarrow} \quad [\bigcirc\bigcirc] \quad \overset{\rm Sh}{{\color{blue}\longrightarrow}} \quad [\bigcirc\bigcirc] \quad \longrightarrow \quad \varnothing
    \label{bip1unknotcomplex}
\end{equation}
with
\begin{equation}\label{Sh-matrix}
    {\rm Sh} =\begin{pmatrix}
        0 & 0 & 0 & 0 \\
        -1 & 0 & 0 & 0 \\
        1 & 0 & 0 & 0 \\
        0 & 1 & -1 & 0
    \end{pmatrix}.
\end{equation}
This new operator $\rm Sh$ possesses the differential properties ${\rm Sh}\,\Delta = 0$ and $m\,{\rm Sh} = 0$ and is consistent with~\eqref{Sh-morph}.

%\newpage

\printbibliography

\end{document}